# Zero-Ending Prices, Cognitive Convenience, and Price Rigidity*

**Avichai Snir[a], Haipeng (Allan) Chen[b], Daniel Levy[a,c,d,e,f] ****

[a] *Department of Economics, Bar-Ilan University, Ramat-Gan 5290002, Israel*
[b] *Gatton College of Business and Economics, University of Kentucky, Lexington, KY 40506, USA*
[c] *Department of Economics, Emory University, Atlanta, GA 30322, USA*
[d] *International School of Economics at Tbilisi State University, Tbilisi, Georgia*
[e] *International Centre for Economic Analysis* (ICEA)
[f] *The Rimini Centre for Economic Analysis* (RCEA)

**Abstract:** We assess the role of cognitive convenience in the popularity and rigidity of 0-ending prices in convenience settings. Studies show that 0-ending prices are common at convenience stores because of the transaction convenience that 0-ending prices offer. Using a large store-level retail CPI data, we find that 0-ending prices are popular and rigid at convenience stores even when they offer little transaction convenience. We corroborate these findings with two large retail scanner price datasets from Dominick's and Nielsen. In the Dominick's data, we find that there are more 0-endings in the prices of the items in the front-end candies category than in any other category, even though these prices have no effect on the convenience of the consumers' check-out transaction. In addition, in both Dominick's and Nielsen's datasets, we find that 0-ending prices have a positive effect on demand. Ruling out consumer antagonism and retailers' use of heuristics in pricing, we conclude that 0-ending prices are popular and rigid, and that they increase demand at convenience settings, not only for their transaction convenience, but also for the cognitive convenience they offer.

**JEL Classification**: E31, L16, D90, E70, M30

**Keywords**: Cognitive Convenience, Transaction Convenience, Price Rigidity, Price Stickiness, Sticky Prices, Rigid Prices, 0-Ending Prices, Round Prices, Convenient Prices, 9-Ending Prices, Just Below Prices, Psychological Prices, Price Points

* We are grateful to two anonymous reviewers for detailed and constructive comments, and to the associate editor for comments, suggestions, and guidance, which helped improve the paper substantially. We thank the participants of the 2014 LABSI conference in Siena, the 2017 Pricing Research Workshop at the University of Illinois, and the 2016 Israeli Economic Association annual conference, particularly Dhruv Grewal, Yishay Maoz, Kent Monroe, Robert Schindler, and Michael Tsiros for thoughtful comments. We are grateful to Ed Knotek for useful conversations and suggestions, and for answering our questions on measuring and quantifying the convenience of various denomination coins. We are particularly grateful to Doron Sayag for providing us with the CPI price and the consumer expenditure distribution data for Israel. An earlier version of this paper was written when the first author was a senior lecturer at Netanya Academic College. All authors contributed equally. The usual disclaimer applies.

** Corresponding author at: Department of Economics, Bar-Ilan University, Ramat-Gan 5290002, Israel
*E-mail address:* Daniel.Levy@biu.ac.il (D. Levy)

Bruno was still standing at the window…
"Don't imperrupt!" he said as we came in. "I'm counting the Pigs in the field!"
"How many are there?" I enquired. "About a thousand and four," said Bruno.
"You mean 'about a thousand,'" Sylvie corrected him. "There's no good saying *'and four'*: you *can't* be sure about the four!"
"And you're as wrong as ever!" Bruno exclaimed triumphantly. "It's just the *four* I *can* be sure about; 'cause they're here, grubbling under the window! It's the *thousand* I isn't pruffickly sure about!"

**Lewis Carroll, *Sylvie and Bruno Concluded*, 1893, Chapter 5, pp. 77–78** (spelling as in the original)

At the museum of natural history, a visitor asks the curator: "How old is this dinosaur over here?"
"Seventy million and thirty-seven years" is the answer.
As the visitor marvels at the accuracy of the dating, the curator explains:
"I've been working here for 37 years, you know, and when I arrived, I was told that it was 70 million years old!"

**Stanislas Dehaene, *the Number Sense: How the Mind Creates Mathematics*, 1997, p. 108**

On January 3, 2012, [Starbucks] set the net price of a tall cup of coffee in Manhattan at precisely $2.01, including tax. "I didn't have the penny" … Ms. Schmais wasn't especially irked by the price increase, which comes to 10 cents for a tall cup. But that orphaned penny had her fuming. "It's the stupidity of it," she said, "It's what I'd call 'the annoyance factor.'…It's ridiculous. Why the extra penny? Who has pennies? Didn't anyone think this through? Couldn't they round down or even up? Why leave it at a penny?" … When David Turnbull presented two $1 bills for coffee at the Starbucks on Astor Place, he met the same problem. He wasn't carrying any pennies. "I can't believe it," he said. "Now I need to walk around with pennies? Who could possibly think a price of $2.01 makes sense?"

**Jeff Sommer, "Dear Starbucks: A Penny for Your Thoughts," *New York Times*, January 15, 2012, p. BU3**

## 1. Introduction

Many retail prices end with 9 because retailers believe that 9-ending prices lead to higher demand. 9-ending prices are also more rigid than non 9-ending prices. At convenience stores, however, the most common prices are round, i.e., ending with 0, such as a newspaper for $1.00, a movie ticket for $6.00, etc. At convenience stores, round prices also tend to be more rigid than other prices.

The popularity of round prices at convenience stores is often explained by *transaction convenience* they offer (i.e., the ease of making the payment and the amount of time it takes). Shoppers at convenience stores often purchase one or two items and tend to pay in cash. In such settings, round prices minimize the number of coins needed. In other words, round prices make the purchase process less time-consuming, and thus offer greater transaction convenience.[1]

In this paper, we argue that in addition to transaction convenience, *cognitive convenience* (i.e., the mental and cognitive ease of making a transaction) plays an important role in the popularity of 0-ending prices at convenience stores. 0-ending prices are cognitively more accessible, and thus they offer greater cognitive convenience because the mental and cognitive efforts needed to process information on 0-ending

[1] Round numbers usually end in 0 (Bellos 2015). The greater the number of the right-hand side zeroes, the "rounder" a number is (Johnson et al. 2009, Snir et al. 2021). Convenient prices usually coincide with existing denomination coins/bills. Therefore, a 0-ending price is usually convenient, but not all convenient prices are round. For example, 25¢ is a convenient price as it can be paid with a single 25¢ coin. Sometimes 5-ending prices are also considered round, but less so than 0-ending prices (Schindler and Kirby 1997). In this paper, our focus is on 0-ending prices.

prices are low.[2]

To test our hypothesis, we employ three large datasets. First, we use a retail price dataset from stores Israel. Following Knotek (2011), we classify a store as a *convenience store* if Israel's Central Bureau of Statistics (CBS) flags it as a convenience store, a small grocery, a kiosk, an open market stall, or a specialty store. We find that at convenience stores, 0-ending prices were common and rigid when non-round endings had little effect on transaction convenience. During the study period, the smallest denomination coin in Israel was 10 Agora (NIS 0.10), yet the use of non 0-ending prices, such as prices ending with 5 or 9, was allowed. In cash transactions, when the final price was not a multiple of NIS 0.10, the law required that the final price be rounded to the nearest 0-ending.[3] Therefore, because of the rounding, 0-ending prices were no more transaction-convenient than nearby non 0-ending prices. And yet, we find that 0-ending prices were more common and more rigid than other prices.

Second, we use data from a large US retailer Dominick's, where most prices are 9-ending, to show that in a category where cognitive convenience is likely to be particularly beneficial, 0-ending prices are common and rigid, and have a positive effect on demand. The US data thus provide convergent evidence that the popularity of 0-ending prices is not driven only by transaction convenience. After all, supermarket shoppers typically buy a basket of goods, and thus the end digit of a single good's price (or a small number of goods' prices) should have little effect on the roundness of the final total price.

Third, we use retail price data from the Nielsen dataset, covering 30,803–47,136 retailers across the US. Despite some important shortcomings because of the way prices are measured and reported in Nielsen dataset, we use it to test the robustness of the results.

The work that is the closest to ours is Wieseke et al. (2016), who also suggest that 0-ending prices facilitate purchases at convenience stores because of the high cognitive accessibility of 0-ending prices. Their empirical evidence, however, is limited to a single product at three university cafeterias, with a few hundred dollars of sales. More important, however, they cannot distinguish the cognitive convenience of 0-ending prices from their transaction convenience. In contrast, we study three large retail price data sets for hundreds of goods from two countries. In addition, the legal setting in place in Israel during our study period creates an environment that helps us rule out transaction convenience as a possible explanation of our results concerning the prevalence and rigidity of 0-ending prices.

---

[2] In his 2017 autobiography *One Buck at a Time*, Macon Brock, the founder of Dollar Tree, a dollar store chain operating about 8,000 stores across the US, explains the benefits that shoppers get form the chain's round, $1.00 prices: "When a customer walked into our store, *she could shut off her brain. She didn't have to think, didn't have to calculate how much she was spending*" (our emphasis). See: https://abcnews.go.com/Business/wireStory/dollar-tree-makes-official-items-now-cost-125-81361316, accessed February 22, 2022.

[3] For example, no change was given if one used a NIS 10 coin to pay any price in the range NIS 9.95–NIS 10.04.

In addition, we explore a possible effect of the 2011 "cottage cheese protests" (Hendel et al. 2017) on the use of 0-ending prices at Israeli stores. Cottage cheese is a staple food in Israel and its price was regulated by the government until 2006. Once the regulation was lifted, the price of cottage cheese increased quickly. In the summer of 2011, a group of activists organized public protests, which escalated quickly.[4] In the peak, about 1 million Israelis (about 12.8% of the population at the time) went to the streets to demonstrate against the high costs of living.[5] Among other things, the protests drew the media's attention to retailers' practices that were deemed "unfair" (Hendel et al. 2017). That could have led retailers to reduce the number of 9-ending prices and increase the number of 0-ending prices, in an attempt to reduce consumer antagonism (Blinder et al. 1998). However, our analysis rules out consumer antagonism, as well as retailers' use of heuristics in setting the prices (DellaVigna and Gentzkow 2019, Strulov-Shlain 2021, Huang et al. 2022), as possible explanations of our results.

Our empirical strategy is to use the three datasets, the Israeli CPI retail price data, the US scanner price data, and the Nielsen data, and four types of analyses, as follows. First, we demonstrate the popularity of 0-ending prices at convenience stores (Israel) and supermarkets (US) by examining the prevalence of 0-ending prices. Second, to demonstrate the rigidity of 0-ending prices, we estimate baseline linear probability models, which are expanded to include various control variables. Third, we assess the effect of 0-ending prices on demand. Fourth, we subject our findings to multiple robustness tests and find that the results we report are robust.

Our finding that 0-ending prices are more rigid than other prices in settings where cognitive convenience is beneficial adds to a growing literature on price rigidity. Price rigidity (or price stickiness) is a phenomenon where prices do not respond to changes in market conditions, or when their response is sluggish and incomplete. In such situations, monetary policy actions might not immediately translate into prices, and thus they can have a real effect. Price rigidity, therefore, is a critical ingredient in modern New Keynesian macroeconomic models because of its implications for monetary neutrality.

[4] See, for example, https://www.ynet.co.il/articles/0,7340,L-4446809,00.html (in Hebrew), accessed February 22, 2022.
[5] Source of the data on Israel's population in 2011: Israel's Central Bureau of Statistics.

The finding that 0-ending prices enhance sales is also related to optimal price information provision in retail settings. Some recent studies conclude that retailers often use prices to conceal information (Gabaix and Laibson 2006, Chen et al 2008, Chakraborty et al 2015, McShane et al 2016, Levy et al 2020, Snir and Levy 2021). Our findings imply that in settings where shoppers make spontaneous decisions on a small number of goods, reducing the cognitive load that is needed to make decisions can increase the likelihood of a sale. In such settings, transparency of price information can benefit both the sellers and the buyers.

For the remainder of the article, we proceed as follows. In section 2, we review the relevant literature. In section 3, we state our formal hypotheses. In section 4, we study the data from Israel. In section 5, we study the data from the US. In section 6, we briefly summarize the results of the robustness tests. We conclude in section 7.

## 2. Literature review

### *2.1. Price rigidity in New Keynesian macroeconomics*

"A central question in macroeconomics is whether nominal rigidities are important" (Eichenbaum et al. 2011, p. 234). Therefore, a large stream of empirical literature on price rigidity tries to assess the extent of price rigidity using a variety of micro-level data. Examples of earlier studies include Carlton (1986) who studies the rigidity of wholesale prices, Cecchetti (1986) who studies the rigidity of magazine prices, Kashyap (1995) who studies mail-order catalog prices, Lach and Tsiddon (1992, 1996, 2007) who study data from a high-inflation period, Levy et al. (1997) who study price rigidity using data from five large US supermarket chains, Blinder et al. (1991, 1998) who survey price adjustment practices at 200 large US firms, and Dutta et al. (2002) and Levy et al. (2002) who study price rigidity using retail orange juice price data from a large US supermarket chain and the corresponding commodity spot price data from the New York Cotton Exchange.

More recent examples include Zbaracki et al. (2004) who employ an ethnographic interview method to study the price rigidity at a large US industrial manufacturer, Bils and Klenow (2004), Klenow and Kryvtsov (2008), and Nakamura and Steinsson (2008) who use micro-level price data that is collected and used by the BLS to construct the CPI, Kehoe and Midrigan (2015) who focus on differences between the rigidity of regular prices and the rigidity of sale prices, and Anderson et al. (2015) who study retail price data from a large US retail chain.

These studies use different data sets from different countries, settings, sectors, markets, and levels of aggregation. Yet, they all share one key finding: Retail and wholesale prices tend to be far more rigid than what the existing price-setting models predict. For recent surveys, see Klenow and Malin (2011), Leahy (2011), and Nakamura and Steinsson (2013).

*2.2. The prevalence and rigidity of 9-ending prices*

There is evidence that price endings can carry connotations about various products or a retailer's attributes (Friedman 1967, Schindler 1984, 1991, 2006, Basu 1997, Ruffle and Shtudiner 2006). Consequently, price endings are not distributed uniformly. For example, studies that use store-level retail price data offer overwhelming evidence that 9-ending prices comprise as much as 30%–95% of retail prices, far higher than the 10% predicted by uniform distribution (Snir and Levy 2021). Earlier evidence comes from small, often hand-collected datasets, that cover a limited number of products, and/or over short time horizons (e.g., Schindler 1984). More recent studies use large scanner datasets covering thousands of products over long time periods (e.g., Snir and Levy 2021).

In a sample of meat product prices in 70 US cities, Twedt (1965) finds that 64% of the prices are 9-ending. Kreul (1982) finds that 58% of the restaurant menu item prices of low-cost products (less than $7) are 9-ending. Huston and Kamdar (1996) report that 45.6% of clothing prices end with 9. Schindler and Kirby (1997) report that 30.7% of consumer goods prices are 9-ending. Stiving and Winer (1997) find that 50.5% and 36.1% of tuna and yogurt prices in their data are 9-ending, respectively.

Lee et al. (2009) study a dataset with more than 1.5 million daily price observations on electronic products sold by 90 Internet-based retailers, covering over a two-year period, and find that 38.7% of the prices end with 9. Analyzing Dominick's weekly retail scanner price data for 18,037 products that cover 29 product categories in the chain's 93 stores over an 8-year period, Levy et al. (2011) find that 69% of the prices are 9-ending. They also find that 9-ending prices are more rigid than other prices.

Anderson et al. (2015) use a price dataset that comes from a large US retailer. They find that 95% of the prices are 9-ending. They also find that in their data, prices that end with 9 are far more rigid than other prices. Strulov-Shlain (2022) studies data with 375 million observations and finds that 61% of the prices end with 9. DellaVigna and Gentzkow's (2019) use Nielsen-Kilts retail price panel data for 1,365 products and find that 78% of the prices in their data end with 9.[6]

It seems that 9-ending prices are popular because they seem to increase sales. Schindler and Warren (1988) find that subjects are more likely to choose an item from a restaurant menu if its price ends with 9. Kalyanam and Shively (1998) report 9-ending pricing increases sales by 12%–76%. Anderson and Simester (2003) find in a controlled experiment with catalogue prices that shoppers view prices that end with 9 as lower than prices that end with other digits. In their study of electronic goods prices on the internet, Lee et al. (2009) also find that 9-ending prices increase consumer purchases.

Ater and Gerlitz (2017) take advantage of a regulatory change in Israel to show that the correlation between 9-endings and price rigidity is causal. Starting January 2014, the use of non 0-ending prices has been prohibited in Israel. Ater and Gerlitz (2017) find that before the regulatory change, 9-ending prices were less likely to change than round prices, but after the change the probability of a price change is the same regardless of the price ending.

The literature offers two main explanations for the popularity of 9-ending prices, image effect and level effect. According to the image effect, 9-endings signal low, decreased, or discounted prices (Quigley and Notarantonio 2015; Schindler and Kibarian 1996). Schindler (1984) suggests that 9-endings signal that the price has not been raised.

According to the level effect, shoppers perceive 9-ending prices as low because they tend to round prices down. This may be due to the rounding down of 9-ending prices because rounding up is cognitively more demanding (Freling et al. 2010; Gabor and Granger 1964; Schindler and Kirby 1997; Manning and Sprott 2009). Alternatively, it may be because of lower attention to the price's rightmost digits due to L-to-R processing of price information, causing shoppers to interpret 9-ending prices as lower than they actually are (Poltrock and Schwartz 1984; Thomas and Morwitz 2005).

[6] Studies conclude that 9-endings are particularly popular for low priced goods. For more expensive goods, sellers tend to use more round endings in their prices. For example, at fast food places we frequently observe prices like $1.99, $2.49, etc., but at more upscale restaurants, we usually see prices like $24, $18, etc. Stiving (2000) concludes that firms that use high prices to signal quality, are more likely to set those prices at round numbers. Consistent with this conclusion, he offers empirical evidence that firms tend to use round prices for higher quality products. In a meta-study, Freling et al. (2010) show that 9-ending prices affect sales, recall, and perceptions of price, quality, and value, although price and product characteristics seem to moderate these effects.

Snir et al. (2017) survey shoppers in Israel and find that they perceive 9-endings as a signal for low prices. Levy et al. (2020) offer evidence from lab experiments, where participants were shown two numbers and were asked to identify as quickly as possible the larger/smaller of the two. The experiment involved two conditions. In one, the numbers were presented as prices of some goods, while in the second, they were presented as numbers. Levy et al. find that in the number condition, 9-endings did not affect the likelihood of a correct response. In the price condition, however, participants used 9-endings as a signal for low prices, reducing the likelihood of a correct response by about 9%. Using price data from Israel and the US, they also find that 9-ending prices are more rigid than other prices. Interestingly, they find asymmetric 9-ending price rigidity: 9-ending prices are more rigid upward than downward.

### *2.3. The prevalence and rigidity of 0-ending prices*

Another common price ending, especially in settings where convenience is important, is 0. One of the first papers that documented the popularity of round numbers in prices is Jones (1896) that offers a series of dataset examples in which round numbers are overrepresented. Jones invokes convenience as the chief reason for the practice. Similarly, Watkins (1911) alludes to convenience to explain the rigid pricing of a street railway ride at 5¢.

Studies that use EU data around the period of the Euro changeover also offer evidence about the importance of convenient prices. According to Deutsche Bundesbank (2002), for example, after the changeover the two most common movie ticket prices in Germany were €6 and €6.50. Before the changeover, the equivalent figures in DM were also round, DM 11 and DM 12, respectively. Similarly, Hoffman and Kurz-Kim (2006) report that the fraction of round prices in Germany were similar before and after the changeover, even though the conversion rate from DM to Euro is not round (i.e., €1 = DM 1.95583). Similar phenomena are documented for the pre- and post-Euro changeover prices by Cornille and Stragier (2007) in Belgium, and by Glatzer and Rumler (2007) in Austria.

Studies also find that in settings where convenience is important, round prices are more rigid than other prices. Galbraith (1936) notes that cutting the price of "cheap confections" from 5¢ to 4¢ would lead to buyers' inconvenience, linking this observation to the rigidity of such round prices.

Levy and Young (2004, 2021) and Young and Levy (2014) report that the price of a 6½ oz serving of Coca-Cola, from the fountain or in a bottle, was 5¢ for 74 years, from 1886 (the year it was first introduced) until 1959. According to Levy and Young, this extraordinary rigidity of the price of Coca-Cola was in part caused by the desire of the Coca-Cola Company to ensure that Coke's price remained convenient: The price of a nickel (i.e., a single coin) facilitated the drink's purchases, first at pharmacies and other outlets where the drink was sold, and later through vending machines, many of which only accepted nickels.

Kashyap (1995) studied mail-order catalog prices for 12 different products during the 1953–1987 period to explore the link between price-endings and price rigidity. Indeed, he finds that in his data all prices end with either 0 or 5, which he defines as "price points." Moreover, he finds that the prices in his data are rigid in the sense that they tend not to move away from the price points.

According to Knotek (2008, 2011), 0-ending prices are designed to simplify and expedite transactions by requiring little or no change. He hypothesizes that shoppers at convenience stores often purchase one or two items, and they tend to pay in cash. In such settings, *convenient* prices minimize the number of coins needed to make a transaction. Indeed, Knotek (2008, 2011) shows that the popularity and rigidity of convenient prices are correlated with a measure of "*transaction-inconvenience*" of the prices, which he defines as the minimum number of coins and notes needed to pay.

Knotek (2008) reports findings for US newspaper prices, while Knotek (2011) considers the prices of other goods, including movie tickets, coffee, public transportation, single-serving drinks, candies, and other convenience-type goods in the US. Fisher and Konieczny (2000, 2006) also note the importance of convenient prices for the rigidity of Canadian newspaper prices.

Bils and Klenow (2004), using a dataset which covers about 70% of the CPI categories, also note the correlation between convenience and price rigidity. They find that consumption spending categories that are most likely to be affected by convenience, such as parking fees, prices at laundromats, prices at vending machines, etc., tend to be among the goods and services with the stickiest prices (Knotek 2008).

Bouhdaoui et al. (2014) conduct a purchase behavior survey among 1,106 French shoppers to demonstrate the link between convenient prices and the common use of cash in transactions at convenience stores and locations. They find that the share of cash payments increases with convenient prices, suggesting that the rigidity of convenient prices can be explained in part by the frequent use of cash to pay convenient prices.

Chen et al. (2019) and Shy (2020) also find evidence that shoppers care about transaction convenience. They explore the link between currency denomination and the use of cash vs. credit cards, and find that the "burden of receiving and carrying change" increases as the transaction value moves away from round prices, such as $5, $10, $20, etc. They find that the peaks of cash use correspond to transaction values that are multiples of $20, the most common denomination dollar bill in ATM machines.

Teber and Çevik (2022) use survey data collected in 2020 by the Central Bank of Turkey on shoppers' preferred method of payment and confirm the findings of Chen et al. (2019) and Shy (2020). They report that the likelihood of cash usage increases with convenient prices and decreases with the transaction amount and the number of coins received in change. They also find that cash usage increases with prices that match currency denominations and conclude that convenient prices increase cash usage.

However, some authors argue that in addition to transaction convenience, 0-ending prices may be common and rigid also for their cognitive convenience. Indeed, the shape of 0 satisfies *Gestalt* principles (Gestalt, a "form" in German, refers to characteristics of visually perceived images): It is simple, balanced, unified and coherent, making it easily accessible to cognitive processes (Treisman and Gelade 1980, Dehaene 1997, Schindler and Kirby 1997, and Palmer 1999).

Chenavaz et al. (2018) show that 0-ending prices are common and rigid in online markets, although online transactions are processed electronically. They explain their findings by assuming that shoppers are less attentive when shopping online compared to shopping in brick-and-mortar shops, making the cognitively accessible round prices more appealing. They reason that people shop more frequently offline than online, and thus they develop their shopping habits offline. For example, people purchase food necessities offline, where they may consciously pay attention to small changes. Therefore, Chenavaz et al. argue, convenient pricing exerts influences in online shops, even though it comes from offline shops.

Schindler and Kirby (1997) argue that because it is easier to process round numbers, retailers can benefit from setting 0-ending prices if they want the shoppers to remember and recall the goods' prices easily.[7]

Wadhwa and Zhang (2012, 2015) suggest that the cognitive convenience of round prices is beneficial also because it makes consumers "feel right" when purchase decisions are emotionally driven.

Isaac et al. (2021) show that a characteristic of the number representing a debt amount impacts the speed of debt repayment. In a series of field studies and experiments, they find that people are likely to pay off round-number debts (e.g., moderate-size debt amounts that end in 0 or 5) more quickly than debts of similar magnitude that are not round numbers. They show that this effect arises because of the cognitive ease of round numbers, which manifests itself in processing-fluency and retrieval-fluency. According to Isaac et al. (2014, p. 242), "…we would expect more mental occurrences of the number 30 than the number 31 because '31' will tend to evoke the thought 'one more than 30,' but '30' will not tend to evoke the thought 'one less than 31' (Isaac and Schindler, 2014). Consistent with this, round numbers have been found to occur in written and spoken language far more frequently than sharp [i.e., non-round] numbers."

Wieseke et al. (2016) conducted a field experiment, where they offered chilled beverages for sale at three cafeterias for either $0.99, $1.00, or $1.01. They find that sales were highest when the price was $1.00, suggesting that at convenience outlets, round prices enhance sales.

## 3. Our hypotheses

[7] Indeed, 0-ending numbers play a special role in our lives. We use them in setting alarm clocks, meeting times, work hours, etc. Round number birthdays or anniversaries are particularly important. People use round numbers when they guess/estimate (Plug 1977). Baseball players try to reach a batting average above 0.30 (Pope and Simonsohn 2011). Drivers at self-service gas stations try to stop the pump at a round price (*The Perfect Pump*, with Jerry Seinfeld, at www.youtube.com/watch?v=m3JVr5HoeoA, accessed February 22, 2022), and there are discontinuous drops in used car sale prices at 10,000-mile (US) and at 10,000-km (Canada) odometer thresholds (Lacetera et al. 2012). Customers at restaurants leave round tips or round up the total bill (Lynn et al. 2013). Kandel et al. (2001) document the significance of 0-endings in IPO auctions in Israel. See also Sonnemans (2006). In Google search (May 4, 2021), the numbers 99, 100, 101, 999, 1000, and 1001 appeared $7{,}720\times10^6$, $25{,}270\times10^6$, $3{,}780\times10^6$, $1{,}420\times10^6$, $9{,}160\times10^6$, and $598\times10^6$ times, respectively. Thus, 0-ending numbers occur in Google more often than nearby non 0-ending numbers.

Following the studies surveyed above, we pose the following formal hypotheses. For convenience stores and conveniently purchased products, even after controlling for the effect of 0-ending prices on transaction convenience, we expect that compared to other prices:

**H1: 0-ending prices are more prevalent.**

**H2: 0-ending prices are more rigid.**

**H3: 0-ending prices increase demand.**

Below, we use retail price data collected by Israel's *Central Bureau of Statistics* (*CBS*) for compiling the CPI, and two retail price datasets from the US to test these three hypotheses. In the analyses, we control for the effect of prices' transaction convenience by (1) focusing on an environment where 0-endings have only a small effect on the transaction convenience, and (2) statistically controlling for the potential effects of transaction convenience. We find that consistent with our hypotheses, 0-ending prices are common, rigid, and that they are associated with higher sales volumes. We conduct multiple robustness checks and find these results continue to hold. In addition, we are able to rule out rival explanations for our results, including consumer antagonism (Blinder et al. 1998) and retailers' use of heuristics in setting the prices (DellaVigna and Gentzkow 2019, Huang et al. 2022).

## 4. 0-ending prices in Israel

We test H1 and H2 concerning the prevalence and rigidity of 0-ending prices using price data from convenience stores in Israel. Prices in Israel are quoted in *New Israeli Shekels* (NIS).[8] By law, posted prices are the final prices, including VAT and any other relevant tax. The currency denominations are 0.10, 0.50, 1, 2, 5 and 10 NIS coins, and 20, 50, 100, and 200 NIS notes. The NIS 0.01 (1-agora) and the NIS 0.05 (5-agora) coins were eliminated in 1991 and 2008, respectively, for their high costs of production.[9]

---

[8] During our sample period, the average NIS/$ exchange rate was NIS 4.09 per $1.

[9] For example, by 2008 the cost of minting a 5-agora coin was 16-agora. Also, the public was reluctant to accept them as change, and vending machines, parking meters, and other coin-operated devices stopped accepting them. See: www.boi.org.il/en/NewsAndPublications/PressReleases/Pages/070716e.aspx, and www.boi.org.il/press/eng/100815/100815d.htm, both accessed February 22, 2022. Levy, et al. (2011, p. 1429) discuss similar episodes of elimination of low-denomination coins in other countries.

Although the NIS 0.01 and NIS 0.05 coins have not been in use since then, until the end of 2013 retailers were free to use any price ending. If a consumer used a credit-card, she paid the exact amount. If she paid with cash, however, the total bill was rounded. Until January 2008, prices were rounded to the nearest NIS 0.05. Since 2008, prices have been rounded to the nearest NIS 0.10. Therefore, in the case of a cash transaction in Israel, unlike in the US, a price such as NIS 9.99 did not require dealing with a change any more than a round price of NIS 10.00. Because of the rounding rule, 0-ending prices offered little transaction convenience advantage over nearby non-0-ending prices. For example, since 2008, for all prices in the range NIS 4.95–NIS 5.04, no change should have been involved if one used NIS 5.00 coin.

Thus, if 0-endings were used for their transaction convenience, then at convenience stores in Israel, 0-ending prices would be no more common than other prices. Actually, to the extent that 9-ending prices offer psychological benefits relative to other prices (Levy et al. 2011, Snir and Levy 2021), one would expect that convenience store operators in Israel would set prices at 9 endings to take advantage of these benefits.

Indeed, if consumers perceive 9-ending prices as low, then 9-ending prices should have been more popular in Israel than in the US. In the US, when a retailer sets a 9-ending price, it loses 1¢ relative to the nearby round price. However, because of the rounding law in Israel, retailers could signal that their prices are low by setting a 9-ending price, without increasing the amount of time required to make a transaction or sacrificing their revenues.

One could argue that perhaps retailers were worried that setting a seemingly low price (9-ending) while actually charging a 1-agora higher price (because of rounding) would antagonize consumers. In other words, a possible explanation could be that retailers at convenience stores use 0-ending prices because they do not want to antagonize consumers by effectively charging them one cent more than the posted price, because shoppers could interpret such pricing as unfair. However, as detailed below in section 4.3.5, we find no evidence of such an effect. Thus, retailers could set a price of NIS 9.99 which shoppers perceive as lower than NIS 10.00, without affecting the revenue or transaction time, because the shoppers that pay in cash would still pay with one NIS 10.00 coin and receive no change. However, we expect that 0-ending prices are more prevalent and more rigid even in this environment (i.e., H1 and H2).

*4.1. Data description*

The dataset contains the prices of individual goods that were collected by the Israeli *Central Bureau of Statistics* (*CBS*) for compiling the CPI. The data cover consumer goods in 125 categories (i.e., Entry Level Items), for January 1999–December 2013. In total, we have 564,742 monthly price observations, 442,811 (78.4%) from convenience stores, and 121,931 (21.6%) from large superstores.

For each store, we have information on the type of store, the city where the store is located, the share of men/women in the city, the city's socio-economic status index, the city's population, the share of minority groups in the population, and the distance of the city from Tel-Aviv.[10] The average price in the sample is NIS 108.4 with a standard deviation of NIS 572.6. The median price is NIS 6.99.

*4.2. Prevalence of 0-ending prices in Israel*

To test the prevalence of 0-ending prices in convenience stores (i.e., H1), Figure 1 shows for January 1999–December 2013, the frequency distribution of price endings in superstores (Panel A) and convenience stores (Panel B). In superstores, which include supermarkets, chain stores, department stores and drugstores, the most common price endings are 9 (54.1%), 0 (33.4%), and 5 (9.5%). The distribution in convenience stores is starkly different: 74.3% of the prices end with 0, 17.5% with 9, and 4.4% with 5.

This emphasizes the difference between convenience stores and superstores. The most common price ending in superstores is 9, perhaps because Israeli shoppers view 9-endings as a signal for low-prices (Snir et al. 2017). Despite this, and even though 0-ending prices can be paid with the same number of coins as the nearby 9-ending prices, in convenience stores the most common price ending is 0. These results support our H1.

However, our data spans 15 years (1999–2013). We therefore ask whether there were any changes in the distribution of 0- and 9-ending prices over time. Figure 2 depicts the shares of 0- and 9-ending prices in Israeli *superstores* (Panel A) and *convenience stores* (Panel B) from 1999 to 2013.

In Panel A, it can be observed that until the 2000s, 9-ending prices were not common in Israel. Until 2003, the share of 9-ending prices was about 20%–30%. After 2005, the share has jumped to about 60%–70%, while the share of 0-ending prices dropped from about 40% to less than 20%. Thus, by 2005, 9-ending prices became as common in Israeli supermarkets as in the US (Levy et al. 2011), suggesting that whatever psychological effects they have on US consumers, similar effects may be present in Israel as well. Indeed, there is evidence that Israeli consumers perceive 9-endings as a signal for low prices (Snir et al. 2017, Levy et al. 2020). After January 2014, the share of 9-ending prices went to zero because of the government's decision to outlaw the use of non 0-ending prices.[11]

[10] Socio-economic status index ranks cities. See https://www.cbs.gov.il/he/publications/doclib/2019/1765_socio_economic_2015/t01.pdf for an example of the data, and https://www.cbs.gov.il/en/publications/Pages/2019/Characterization-and-Classification-of-Geographical-Units-by-the-Socio-Economic-Level-of-the-Population-2015.aspx, for the methodology, both accessed February 22, 2022.

Panel B suggests that the share of 0-ending prices in Israel at convenience stores was about 80% in 1999. It dropped slightly in the 2000s but remained at or above 70% throughout most of the sample period. Thus, the share of convenient prices in Israel was at least as high, and even higher, than what Knotek (2011) reports for the US.

The figure shows that there was a decrease in the share of 9-ending prices and an increase in the share of 0-ending prices at both superstores and convenience stores after 2011. This could be due to retailers' reaction to the consumers' "cottage cheese protests" (Hendel et al. 2017). Below, we use the timing of the consumer protests in one of the robustness tests of our results.

Recent studies suggest that retailers might not set prices optimally, perhaps because of managerial inertia or because it takes time to learn about market fundamentals (DellaVigna and Gentzkow, 2019, Strulov-Shlain, 2021, Huang et al., 2022). It is possible, therefore, that the share of 0-ending prices at convenience stores remained high not because of the optimality of 0-ending prices, but because managers failed to appreciate the advantage of switching to 9-ending prices.

However, there are reasons to believe that this is unlikely in our case. First, Strulov-Shlain (2021) and Huang et al. (2022) show that retailers need about 4–6 quarters to set optimal prices. In Israel, convenient stores had at least 10 years (2003–2013) to learn about the advantages of 9-ending prices. The fact that the share of 0-ending prices remained at around 70% even after such a long period suggests that convenience stores' managers likely had reasons other than inertia for not setting more 9-ending prices.

Second, there was a relatively sharp increase in the share of 9-ending prices and a decrease in the share of 0-ending prices between 2003 and 2006. After 2006, however, the share of 0-ending prices stabilized.

[11] Non 0-ending prices still exist for 26 basic food items whose prices are controlled by the government.

Third, following the 2011 “cottage cheese protests,” there was a large increase (decrease) in the share of 0-ending (9-ending) prices, indicating that convenience stores have a capacity to respond relatively quickly to changes in consumers’ tastes. It is unlikely that they responded strongly when they felt that 9-ending prices were out of vogue but failed to respond when 9-ending prices had a positive effect on demand.

### *4.3. Rigidity of 0-ending prices*

Next, we test H2. If in Israeli convenience stores 0-ending prices are more popular than other prices for their positive effect on demand, then 0-ending prices will also tend to be more rigid than other prices. As Knotek (2008, 2011) notes, if a price-ending has a positive effect on demand, retailers will tend to change it only when the new price has the same ending. This will result in a 0-ending prices changing less often than other prices.

Figure 3 provides a visual demonstration of the rigidity of 0-ending prices at convenience stores and of 9-ending prices at superstores. In each panel, the *x*-axis gives the average percentage of the 0-ending or 9-ending prices that change in a given week. The *y*-axis gives the average percentage of non 0-ending or non 9-ending prices that change in a given week. On each panel, the 125 dots represent 125 product categories.[12] The diagonal lines form 45°. Dots lying below the diagonal line indicate that the *x*-variables are more rigid, i.e., less likely to change, than *y*-variable in the corresponding category. Dots lying above the diagonal lines indicate the reverse.

The top left panel depicts the data of 9-ending and non 9-ending prices at superstores: 9-ending prices are more rigid than other prices in 76.4% of the product categories. The bottom right panel depicts the data of 0-ending and non 0-ending prices at convenience stores: 0-ending prices are more rigid than other prices in 81.6% of the product categories. The top right panel shows that 0-endings prices are less rigid than other prices at superstores, while the bottom left panel shows that 9-ending prices are less rigid than other prices at convenience stores.[13] Thus, Figure 3 suggests that in Israel, similar to the

[12] We exclude product categories with fewer than 50 observations of non 0-ending or non 9-ending prices.
[13] Note that when we compare 0-ending prices to other prices, the latter include 9-ending prices. Similarly, when we compare 9-ending prices to other prices, the latter include 0-ending prices.

US (Levy et al., 2011, Knotek, 2011), 9-ending prices are more rigid than all other prices at superstores, and 0-ending prices are more rigid than all other prices at convenience stores. These results provide preliminary evidence in support of H2.

*4.3.1. Baseline regression results*

As a formal test, we estimate a series of linear probability models. The dependent variable is a dummy variable that equals 1 if the price of good $i$ in store $j$ has changed between month $t-1$ and month $t$, and 0 otherwise. The main independent variables are a dummy for 0-ending prices that equals 1 if the right-most digit of the previous price is 0, a dummy for 9-ending prices that equals 1 if the right-most digit of the previous price is 9, and a transaction inconvenience score, which we define as the minimum number of coins/notes necessary for paying the price.[14] The higher the transaction inconvenience score, the more coins one needs to handle, which increases the amount of time and the number of computations needed to complete a cash transaction. The estimation results are reported in Table 1.

In column (1), we present the results of the baseline regression which also includes fixed effects for product categories, districts,[15] stores, and the quarter to which an observation pertains to. We cluster the standard errors at the level of the stores where the corresponding observations were collected.

The coefficients of the dummies for 0-ending ($\beta = -0.072, p < 0.01$) and for 9-ending ($\beta = -0.036, p < 0.01$) prices are both negative and statistically significant. Therefore, a 0-ending price is 7.2% less likely to change, and a 9-ending price is 3.6% less likely to change, than other prices. Compared to the overall likelihood of a price change in our data (i.e., 38.3%), 0-ending and 9-ending prices are 18.8% and 9.4% less likely to change than the average price, correspondingly. These results provide empirical

[14] Following Knotek (2008, 2011) and Chenavaz et al (2018), the inconvenience score equals the smaller of the two scores obtained using (1) a greedy algorithm, and (2) an indirect greedy algorithm. The former calculates the minimum number of coins/bills needed to pay the exact price. The latter assumes that the shopper pays with the smallest coin/bill that exceeds the price and calculates the smallest number of coins/bills necessary to pay the change. For example, the minimum number of coins needed to pay a price of NIS 6.30 is 5 according to the greedy algorithm (a shopper pays one coin of NIS 5, one coin of NIS 1 and three coins of NIS 0.10). The minimum number of coins needed to pay the same price according to the indirect greedy algorithm is 6 (the shopper pays one NIS 10 coin and receives from the seller as change one NIS 2 coin, one NIS 1 coin, one NIS 0.50 coin, and two NIS 0.10 coins). The inconvenience score is therefore 5. In calculating the number of coins needed to pay a price, we consider the fact that in December 2007 the Bank of Israel introduced the NIS 2 coin, and that as of January 1, 2008, the NIS 0.05 coin is no longer in use.

[15] There are 6 districts: Center, Haifa, Jerusalem, North, South, and Tel-Aviv.

support to our H2.

The coefficient of the transaction inconvenience score is positive and significant ($\beta = 0.006, p < 0.01$). Thus, consistent with our hypotheses, (a) 0-ending and 9-ending prices are more rigid than other prices, and (b) less transaction-convenient prices are more likely to change.

To compare the effects of changing a price from non 0-ending (non 9-ending) to 0-ending (9-ending) on price rigidity with the effect of the transaction inconvenience, we ask how many more coins one would need to pay a price to get an effect that is not statistically significantly different from changing the price from non 0-ending (non 9-ending) to 0-ending (9-ending). We find that the reduction in the likelihood of a price change associated with a 0-ending (9-ending) price is equivalent to needing 8 (3) more coins for paying a price. The average number of coins that is needed to pay a price in our data is 3.15 with a standard deviation of 3.07. The effect of 0-ending (9-ending) prices is therefore equivalent to a change of 2.61 (0.98) standard deviations in the number of coins needed to pay a price.

In column (2), we differentiate between convenience stores and superstores by adding a dummy for convenience stores, and its interactions with the dummies for 0-ending prices and for 9-ending prices, and with the transaction inconvenience score. In this setting, the main effects of the 0-ending prices, 9-ending prices, and the transaction inconvenience scores pertain to their effects in superstores, while the coefficients of the interactions capture their effects in convenience stores relative to superstores.

We find that the 0-ending price dummy ($\beta = -0.003, p > 0.83$) is not statistically significant, while the 9-ending price dummy ($\beta = -0.070, p < 0.01$) is negative and statistically significant. Thus, in superstores, 9-ending prices are more rigid than other prices, but 0-ending prices are not. The coefficient of the transaction inconvenience score is positive and statistically significant ($\beta = 0.002, p < 0.04$), implying that an increase in the number of coins needed to pay a price increases the likelihood of a price change.

The interaction of the dummy for convenience store with the 0-ending price dummy is negative and statistically significant ($\beta = -0.057, p < 0.01$), while its interaction with the 9-ending dummy is positive and statistically significant ($\beta = 0.045, p < 0.01$). Adding the interaction effects to the main effects, we find that 9-ending prices are more rigid than other prices also at convenience stores ($F = 8.8, p < 0.01$). However, when we compare 0-ending prices to 9-ending prices at convenience stores, we find that 0-ending prices are more rigid ($F = 26.07, p < 0.01$).

The interaction of convenience stores with the transaction-inconvenience index is positive and statistically significant ($\beta = 0.003, p < 0.05$), implying that the correlation between the number of coins needed to pay a price and the likelihood of a price change is larger at convenience stores than at superstores.

*4.3.2. Regressions with control variables*

In column (3), we add city-level control variables which include the log of the population, the socio-economic score, the share of women, the log of distance from Tel-Aviv, and the share of minority groups. For each product, we also add the price-level defined as the previous week's price rounded to the nearest NIS and a dummy for 5-ending prices. We include the price-level because as prices increase, the distance between 0-ending and 9-ending price points shrinks in relative terms, which may facilitate price changes. We add the 5-ending price dummy because 5-ending prices can also be considered round prices (Schindler and Kirby 1997).

When we add these variables, the coefficient of 0-ending prices at superstores is negative and statistically significant ($\beta = -0.027, p < 0.05$). Other than that, we still find that 9-ending prices are the most rigid prices at superstores, while 0-ending prices are the most rigid at convenience stores.

*4.3.3. Controlling for sale prices*

In column (4), we add a control for sale prices, which we detect using a sales filter. Sales filter is an algorithm that identifies a price as a sale price if the price goes down, stays there for some time, and then goes back up (Nakamura and Steinsson 2008, Knotek 2019, Ray et al. 2021). Following Knotek (2019), we define a price as a sale price if (i) it decreases and then goes back up (not necessarily to the previous level), and (ii) the sale does not last more than one month. We find that prices are more likely to change following a sale ($\beta = 0.441, p < 0.01$), because sale prices tend to bounce back to regular prices. The coefficients of 0-ending prices, 9-endings prices, transaction inconvenience and their interactions with the dummy for convenience stores remain almost unaffected.

*4.3.4. Small vs. big ticket items*

In columns (5) and (6), we test whether the rigidity of 0-ending prices varies between small and big-ticket items. Levy et al. (2011) show that in the US, 9-ending prices are rigid at superstores for both small-ticket and big-ticket items. Knotek et al. (2021) report similar findings for 0-ending prices at convenience stores.

To explore this, we split the sample into two sub-samples. The first (column 5) includes all store-product combination with a maximum price of up to NIS10. The second (column 6) includes all other store-product combinations. We choose NIS10 as the threshold because this is the value of the largest coin denomination – i.e., the highest price that can be paid using a single coin.

We find that in the first sub-sample, which includes about 45% of the observations, the coefficient of 0-ending price dummy is positive ($\beta = 0.073, p < 0.06$) while the coefficient of 9-ending price dummy is negative ($\beta = -0.067, p < 0.01$). Thus, we find that at superstores, 9-ending prices are more rigid and 0-ending prices are less rigid than other prices.

The coefficients of the interaction of the dummy for convenience stores with the 0-ending price dummy is negative ($\beta = -0.122, p < 0.01$), while the interaction with 9-ending price dummy is positive ($\beta = 0.041, p < 0.05$). Adding the interaction and main effects for 0- and 9-ending prices, we find that at convenience stores both 0-ending ($F$ = 36.6, $p < 0.01$) and 9-ending prices ($F$ = 7.6, $p < 0.01$) are more rigid than other prices, with 0-ending prices being more rigid than 9-ending prices ($F$ = 8.15, $p < 0.01$).

For prices larger than NIS10, we find that both 0-ending ($\beta = -0.093, p < 0.01$) and 9-ending prices ($\beta = -0.120, p < 0.01$) are more rigid than other prices at superstores. The difference between the coefficients is statistically significant ($F$ = 8.9, $p < 0.01$). The coefficient of the transaction inconvenience score ($\beta = 0.005, p < 0.01$) is statistically significant but it is smaller than the coefficient obtained in the sample of prices below NIS10, suggesting that adding one additional coin to the number of coins needed to pay a price has a smaller effect on price rigidity when the price is relatively high.

The coefficient of the interaction of the dummy for convenience stores with the dummy for 0-ending prices is negative ($\beta = -0.054, p < 0.01$), implying that 0-ending prices are more rigid in convenience stores than in superstores. The coefficient of the interaction of the dummy for convenience stores with the dummy for 9-ending prices is positive but not statistically significant ($\beta = 0.020, p > 0.12$). Thus, the rigidity of 9-ending prices is not statistically different at convenience stores than at superstores. Nevertheless, at convenience stores, 0-ending prices are more rigid than 9-ending prices ($F = 37.8$, $p < 0.01$).

In summary, when we focus on products costing less than NIS10, we find that at superstores, 9-ending prices are the most rigid prices while 0-ending prices are less rigid than other prices. At convenience stores, 0-ending prices are the most rigid prices, with 9-ending prices also being more rigid than other prices. Focusing on products costing more than NIS10, both 0- and 9-ending prices are more rigid than other prices at both convenience stores and superstores. In addition, transaction convenience seems to play a larger role for products costing less than NIS10 than for products costing more than NIS10. Therefore, our results on the rigidity of 0-ending prices are robust and provide strong support for H2.

#### *4.3.5. Checking for consumer antagonism*

A possible explanation for our findings is that retailers at convenience stores use 0-ending prices because they do not want to antagonize consumers by effectively charging them one cent more than the posted price.[16] But then consumer antagonism should have also become an issue in superstores after 2008, because until 2008, the NIS 0.05 coin was a legal tender. Consequently, total bill with four endings {1, 2, 6, 7} were rounded downwards, and with four endings {3, 4, 8, 9} was rounded upwards. In January 2008, the NIS 0.05 ceased to be a legal tender. Therefore, only four endings were rounded downwards {1, 2, 3, 4} while five endings were rounded upwards {5, 6, 7, 8, 9}. This asymmetry could have made consumers frequenting supermarkets aversive to odd prices, because of the higher likelihood of paying the higher, upward rounded price.[17]

Figure 2, however, suggests that the share of 9-ending prices did not decrease after 2008, not in superstores nor in convenience stores. To explore this further, we test whether there was a change in the rigidity of 0-ending and 9-ending prices following

[16] Convenience stores could be using round prices to discourage shoppers from paying with credit cards. Since paying with credit cards does not entail rounding, a 9-ending price might encourage a shopper who wants to avoid paying an extra cent to use a credit card. This is unlikely, however, because shoppers at convenience stores usually know that the price is higher than at superstores and, therefore, are unlikely to choose paying with a credit card only to save a penny.
[17] The government explained the banning of non-round prices by arguing that in many cases consumers paid a higher price than the one they observed at the till (Ater and Rigby, 2017).

2008 by adding to the regression a dummy for post-2008 (1 for prices after January 2008, 0 otherwise) as well as its interaction with the 0-ending price dummy, 9-ending price dummy, convenience store dummy, and the transaction inconvenience score. The estimation results are reported in column (7).

We find that before 2008, both 0-ending ($\beta = -0.037, p < 0.01$) and 9-ending prices ($\beta = -0.054, p < 0.01$) were more rigid than other prices at superstores. The coefficient of the interaction of the convenience store dummy with 0-ending price dummy is negative ($\beta = -0.045, p < 0.01$), while with 9-ending price dummy it is positive ($\beta = 0.033, p < 0.02$). Thus, 0-ending prices were more rigid at convenience stores than at superstores, while 9-ending prices were more rigid at superstores than at convenience stores.

Focusing on the period after 2008, we find that the interaction of 9-ending price dummy with the dummy for post-2008 is negative ($\beta = -0.046, p < 0.01$). Thus, after 2008, 9-ending prices became more rigid at superstores in comparison to the pre-2008 period. The interaction of convenience stores with the 9-ending price dummy after 2008 is positive but not statistically significant ($\beta = 0.012, p > 0.21$). We therefore conclude that after 2008, 9-ending prices were more rigid than they were before 2008 at both superstores and convenience stores ($F = 11.0, p < 0.01$).

These results suggest that although the elimination of the NIS 0.05 coin could have increased retailers' concern that 9-ending prices might antagonize the consumers, we find no evidence that 9-ending prices were used less or that they became less rigid. We find that after 2008, 9-ending prices actually became more rigid in both superstores and convenience stores in comparison to the pre-2008 period.

As another test for the possible effect of consumer antagonism on retailers' behavior, we include a dummy for the period after the "cottage cheese protests" of the summer of 2011 as well as its interactions with the controls for 0-ending prices, 9-ending prices, and convenience stores, and with the transaction inconvenience score.

The results are reported in column (8). We find that following the protests, the rigidity of 9-ending prices did not change at superstores ($\beta = 0.009, p > 0.54$), nor at convenience stores ($\beta = -0.001, p > 0.90$).

Thus, the data does not support the hypothesis that a reason for using round prices is to minimize consumer antagonism. On the contrary, we find evidence that in superstores, as well as in convenience stores, retailer's preference for 9-ending prices, as manifested by the rigidity of 9-ending prices, has remained unchanged or even increased during periods when such concerns should have been the most acute.

## 5. 0-Ending Prices in the US

We argue above that 0-ending prices are popular not only because they provide transaction convenience, but also because they provide cognitive convenience. In the previous section, we showed that 0-ending prices are more prevalent and more rigid than other prices in an environment where 0-endings have little effect on the transaction convenience (i.e., H1 and H2). In this section, we test H1 and H2 again by using data from a large US supermarket chain Dominick's, and showing that 0-ending prices are common and rigid for items that are placed in the front-end, point-of-purchase displays (i.e., conveniently purchased products). We then test H3 by showing that 0-ending prices are also correlated with a greater demand for these items.

In the supermarket chain we study, 69% of the prices are 9-ending (Levy et al. 2011, Knotek 2019), and shoppers typically buy a basket of goods. Therefore, even if some of the product prices are 0-ending, whether the total final price is round or not is a matter of chance. Thus, the price endings of goods offered on front-end-displays should have little effect on the time needed to pay the final price.

However, in the front-end-display environment, 0-ending prices might offer cognitive convenience. All shoppers pass by the front-end display shelves as they approach the cash registers.[18] Shoppers who buy from these displays usually buy one or two items, and their purchase decision is usually spontaneous.[19] If the items are missing, or if their prices exceed the shoppers' reservation price, then the shoppers proceed to the cash register without buying anything. 0-ending prices can therefore facilitate purchases of these items by making it easier for the shoppers to process the price information and decide whether the price is below their reservation price or not (Wieseke et al., 2016).

---

[18] A National Consumer Agency study reports, for example, that 90% of consumers bought at least once in 6 months a good displayed on the front-end shelves, while waiting at the cash register. Source: http://dechert-hampe.com /images/stories/Candy_and_Snack_Today_-_Missing_Opportunities_in_Grocery.pdf, accessed February 22, 2022.
[19] Indeed, in the retail industry, front-end displays are also known as "impulse display shelves." See, for example, p. 31 in the product list catalogue of CED, an UK-based manufacturer of display cabinets and display shelves, at https://partenaire.tournus.com/CED_pricelist.pdf, accessed February 22, 2022.

### *5.1. US retail scanner price data*

We use data from a large US Supermarket chain, *Dominick's*, that operated in the Chicago metro area. The dataset contains 98,691,750 weekly price observations for 18,037 products in 29 product categories, during an 8-year period, 1989–1997. For a detailed description of the data, see Barsky et al. (2003), Ray et al. (2006), Chen et al. (2008), Levy et al. (2011, 2020), and Peltzman (2000).

In the early part of the sample period, Dominick's has cooperated with the faculty of the University of Chicago Booth School of Business in conducting various price sensitivity and price adjustment experiments (Hoch et al., 1994), which likely affected both the prices and the resulting consumer demand. In our main analysis, we therefore use data from January 1991 onwards. The data cover 93 stores of the chain.[20] The price observations are the actual transaction prices that were paid each week, as recorded by the checkout cash register scanners.

Although the data is somewhat dated, it has three main advantages. First, it contains both the transaction price (i.e., the actual price consumers paid) and the corresponding quantity sold for each week. Second, it contains additional useful information, such as the wholesale price for each product (which we use as an instrumental variable; see section 6.2). Third, the dataset is large, containing over 98 million weekly observations on more than 18 thousand products sold in 93 large stores during an 8-year period. Data sets that contain similar information are rare, and those that exist have some shortcomings. For example, the Nielsen database does not contain information on wholesale prices. Also, the price data it contains are weighted average of weekly prices (i.e., they are not transaction prices; Strulov-Shlain, 2022), making it less suitable for our analyses. That is why Dominick's data is still employed to study pricing behavior (Midrigan 2011; Knotek 2019; Levy et al. 2020; Snir and Levy 2021). We nevertheless make use of the Nielsen data for a robustness test, as discussed in section 5.3 below and in Appendix D.

Front-end, point-of-purchase display shelves are usually used to sell books, magazines, candies, snacks, sodas, razor blades, batteries, etc. Dominick's dataset, however, contains information only on candies, which the database classifies as the front-end-candies category. It contains 3,727,128 weekly price observations for 503 different candy products, including chocolate bars, candy bars, chewing gum products, etc.[21] The share of price changes in the category is 12.5%. The median price is $0.55, the average price is $0.62, and the standard deviation is $0.24.

---

[20] The data can be downloaded from www.chicagobooth.edu/research/kilts/datasets/dominicks, accessed February 22, 2022.

### *5.2. The prevalence of 0-ending prices in the front-end-candies category*

To test H1, we compare the distribution of price endings in the front-end-candies category and in the remaining 28 product categories. As shown in Figure 4, there are two stark differences. First, the share of 9-ending prices in the front-end-candies category is 39.0%, compared to 65.0% in the other 28 categories.

Second, the shares of the two convenient price endings, 0 and 5, are higher in the front-end-candies category than in the other 28 categories. In particular, the share of 0-ending prices in the front-end-candies category (i.e., 24.1%) is the highest among the 29 product categories. It is more than 6 times higher than the average share of 0-ending prices in the other 28 categories (i.e., 3.8%). The frozen-dinners category has the next highest share of 0-ending prices with 13.3%, followed by the cigarettes' category with 8.7%. In the remaining 26 categories, the share of 0-ending prices is less than 5%. These results offer a preliminary support for H1, in that 0-ending prices are more prevalent than other prices except for 9-ending prices.

### *5.3. The rigidity of 0-ending prices in the front-end-candies category*

To test H2, Figure 5 depicts the percentage of price changes of 0-ending, 9-ending, and all other prices at Dominick's, in the front-end-candies category. The likelihood of a price change for other prices (i.e., 15.96%) is more than twice the likelihood that a 9-ending price will change (i.e., 7.16%), and 76% higher than the likelihood that a 0-ending price will change (i.e., 9.06%).

#### *5.3.1 Baseline regression results*

As a formal test, Table 2 reports the results of estimating regressions of the probability of a price change using a set of linear probability models. The dependent variable in all regressions is a dummy that equals 1 if the price of good *i* in store *j* has changed between week $t-1$ and week $t$, and 0 otherwise. All the independent variables are defined as before. In column 1, the regression also includes fixed effects for weeks and for sub-categories×stores×weeks, i.e., sub-categories in a store, in a given week. We cluster the standard errors at the store level.

[21] For the full list of the products, see Dominick's Data User Manual, pp. 287–298, available at research.chicagobooth.edu/~/media/5F29F56C65894FA194132DB8D36292B3.pdf, accessed February 22, 2022.

We find that 0-ending prices are less likely to change than other prices ($\beta = -0.021, p < 0.01$), although not to the same extent as 9-ending prices ($\beta = -0.137, p < 0.01$). Transaction inconvenience increases the likelihood of a price change ($\beta = 0.045, p < 0.01$).

Although the coefficient of the 0-ending price dummy is not as large as that of the 9-ending price dummy, it is still economically significant: compared to the likelihood that a price that is not 0-ending nor 9-ending will change (i.e., 15.96%), we find that 0-ending and 9-ending prices are 13.2% and 82.7% less likely to change, respectively. These results provide partial support to H2, in that 0-ending prices are more rigid than other prices except for 9-ending prices.

*5.3.2. Regressions with control variables*

To check the robustness of the above results, in column (2) we add the following control variables: a dummy for prices that end in either 0.25 or 0.75 (because such prices can be paid using quarter-coins), a control for the price-level defined as above, a control for the percentage change in the wholesale price, a dummy for 5-ending prices, and a dummy for sale prices in the previous week.[22]

The results still suggest that 9-ending prices are the most rigid ($\beta = -0.156, p < 0.01$), but 0-ending prices are also significantly more rigid ($\beta = -0.104, p < 0.01$) than other prices. We also find that 5-ending prices are less likely to change than other prices ($\beta = -0.141, p < 0.01$).

Thus, in the front-end-candies category, there are three endings that are particularly rigid. The first is 9, which is common and rigid in all product categories at Dominick's (Levy et al., 2011, 2020). The other two are "round" endings, 0 and 5, suggesting that in this category, round, convenient prices are particularly rigid.

[22] The dataset contains weekly wholesale prices, which are measured as the average acquisition cost of the items in inventory. We remove 22,239 observations (0.6%) that have wholesale price changes greater than 150%. To identify a sale price, we could use Dominick's sales' indicator variable, which is included in the dataset. However, according to Peltzman (2000), the variable was not set on a regular basis and consequently it is unreliable. We therefore use a sales filter algorithm to identify sale prices.

In column (3), we test whether 0-ending prices are also more rigid than other prices among regular prices. The reason for our attention on regular prices is the finding reported by some studies that regular prices might be more important for inflation than temporary sale prices. See Nakamura and Steinsson (2008), Eichenbaum et al. (2011), Midrigan (2011), Klenow and Malin (2011), Guimaraes and Sheedy (2011), Beradi et al. (2015), Kehoe and Midrigan (2015), Anderson et al. (2017), and Ray et al. (2021).[23] We therefore remove observations on sale prices and on bounce-back prices (i.e., if the price in the previous week was a sale price), to obtain a sample of regular prices, which we use to re-estimate the regression.

We find that among regular prices, the effect of 0-endings is negative and statistically significant ($\beta = -0.067, p < 0.01$). The effect is economically significant as well. The average probability that a non-0 and non-9-ending regular price changes is 8.3%. Therefore, 0-ending prices are 80.7% less likely to change than other prices.

In column (4), we include in the regression only observations on sale prices. Among sale prices, 0-ending prices are significantly more rigid than other prices ($\beta = -0.109, p < 0.01$), consistent with Strulov-Shlain (2022) who finds that 0-endings lead to high demand if a product is on sale.

In sum, we find that although 9 is the most common and the most rigid price ending in Dominick's data overall (Levy et al., 2011, Knotek, 2019), in the front-end-candies category 0-ending prices are also common and more rigid than prices with other endings. The robustness checks therefore provide convergent support to H1 and H2 using a different dataset.

Note that the high frequency of 0-ending prices is unique to the front-end-candies category. In Appendix B, we show that the rigidity of 0-ending prices at Dominick's is limited to the front-end-candies category, suggesting that cognitive convenience is particularly relevant to this category.

*5.4. The effect of 0-ending prices on demand*

We proceed to test H3, i.e., the effect of 0-ending prices on demand for conveniently purchased product. One possible explanation for the popularity and rigidity of 0-ending prices is that Dominick's does not set prices optimally (DellaVigna and Gentzkow, 2019, Huang et al. 2022). This can be either because the retailer mistakenly believes that 0-ending prices facilitate transactions, or because the retailer uses 0-ending prices as a rule of thumb to set prices. This seems unlikely, however, because the dominant price ending at Dominick's is 9. Therefore, if Dominick's were to set prices because they believe in price points, or because they follow a rule of thumb, they would likely stick to the usual 9-ending prices that they are using in other product categories.

[23] For dissenting opinions, see Klenow and Kryvtsov (2008), Fox and Syed (2016), Glandon (2018), and Kryvtsov and Vincent (2021).

Nevertheless, to rule out this possibility, we directly test the effect of 0-ending prices on demand. Dominick's data contains, for each store, the number of units sold each week. For illustration, Figure 6 shows the scatter plot of the weekly number of Breath Savers Wintergreen Sugar Free Mints sold at Store No. 101. The black line depicts the log-log regression line, where the independent variables is the price. The figure illustrates graphically that when the price is round (red circles), the demand is higher than predicted in 70.14% of the weeks.

### *5.4.1. Baseline regression results*

To test the effects of 0-ending prices on demand more formally, we follow Strulov-Shlain (2022) in estimating log-log demand equations with instrumental variables (Hoch et al. 1995, DellaVigna and Gentzkow 2019, Butters et al. 2020). Because we are using scanner data, missing observations could be due to zero sales, which are more likely when the price is high. We therefore focus on more popular products, which we define as products that were available for at least 50% of the weeks in all stores (Strulov-Shlain, 2022). This leaves us with 80 products and 1,988,759 observations.

We thus estimate 80 product-level, reduced-form demand equations, one for each product. The dependent variable in each regression is the log of the number of units of product $q$ sold in store $s$, in week $t$. The independent variables include dummies for 0- and 9-ending prices, the transaction inconvenience score, the log of the price, the log of the average price of other products in the same sub-category in the store as a control for competition, the log of the quantity sold in week $t - 1$ (Hoch et al. 1995), and fixed effects for years and months to control for possible seasonality in demand (Butters et al.

2020), and for stores. We also add fixed effects for holidays (Levy et al. 2010).[24] To minimize the effect of endogeneity, we follow DellaVigna and Gentzkow (2019) and Strulov-Shlain (2022) and use the average price of the products in all other stores as the instrument for the price.

Panel A of Table 3 reports the results. The first column reports the average of the coefficients over the 80 products. The second column reports the results of the quantity-adjusted mean. The third (fourth) column reports the number of positive (negative) coefficients. The fifth (sixth) column reports the number of statistically significant positive (negative) coefficients.

We find that the price coefficients' average is −0.264. Thus, the elasticity of demand in the front-end-candies category is quite small, consistent with Andreyeva et al. (2010) and Hoch et al. (1994) who also find that demand elasticity for candies is low.

More important, 0-ending prices have a positive effect on demand: the mean coefficient of the 0-ending dummy is positive, 0.173. We also find that there are 51 positive coefficients compared to 29 negative ones. If we compare statistically significant coefficients, we find that the number of positive and significant coefficients (i.e., 43) is 3.3 times higher than the number of negative and significant coefficients (i.e., 13). These results support H3 that 0-ending prices increase demand for conveniently purchased products.

9-ending prices do not increase the demand. The average coefficient of 9-ending prices dummy is −0.009, and the number of positive and significant coefficients (i.e., 28) is smaller than the number of negative and significant coefficients (i.e., 38).

*5.4.2. Regressions with control variables*

In panel B of Table 3, we report the results of regressions to which we add a full set of controls: a dummy for sale prices, a dummy for prices that end in 25 or 75 and thus can be paid using quarter-coins, and a dummy for 5-ending prices. The effect of 0-ending prices (i.e., 0.122) remains positive and the number of positive and statistically significant coefficients is 28, compared to 17 negative and statistically significant coefficients. The coefficient of 9-ending prices is still negative (i.e., −0.012), with fewer positive and significant coefficients (i.e., 16) than negative and significant coefficients (i.e., 30).

[24] The holidays include Christmas, New Year, Presidents Day, Easter, Memorial Day, 4th of July, Labor Day, Halloween, and Thanksgiving.

Strulov-Shlain (2022) also finds a positive effect of 0-endings on demand, but he argues that this effect is mostly due to sale prices. As a further test, therefore, in panel C of Table 3 we use only observations on regular prices. This has little effect on our main results: the coefficient of 0-ending prices (i.e., 0.145) is positive. When we focus on regular prices, we find that the coefficient of 9-ending prices (i.e., 0.026) is also positive.

In panel D of Table 3, we report the results of regressions of sale prices. Some products were hardly on sale, and thus we drop all products with less than 1,000 sale price observations, leaving us with 49 products and 140,606 observations. We find that the coefficient of 0-ending prices is positive (i.e., 0.006) but smaller than for regular prices. The coefficient of 9-ending prices is negative (i.e., −0.029). Thus, when we focus of the front-end candy category, we find that the effects of 0-endings are stronger for regular prices than for sale prices. We recognize, however, that the sample size for sale prices is relatively small.

In sum, the robustness checks provide strong evidence that 0-ending prices have a positive effect on demand, supporting H3. This further strengthens our belief that the retailer uses 0-ending prices because they facilitate sales, not because the retailer fails to set prices optimally.

## 6. Robustness tests

We run multiple robustness tests to assess the sensitivity of the results we are reporting. Below we briefly discuss these tests. The discussion is structured around the three datasets we use. The tests are described in more detail in the Online Supplementary Web Appendixes A–D.

### *6.1. Robustness tests with the Israeli CPI data*

In the appendix, we discuss a range of robustness tests we run to further differentiate between transaction convenience and cognitive convenience. In the tests we consider the sub-samples of our data where cognitive convenience might be particularly important such as sub-samples of low prices, and a subsample of prices that can be paid with a single coin but with different price endings. We also differentiate between convenience stores where transaction convenience might be particularly important and other convenience stores, and between periods with different costs of transaction inconvenience. In addition, we check if we lost information by bunching all non-0 and all non-9 ending prices together. Finally, we analyze a subsample of regular prices because regular prices might play a different role in the inflationary process than sale prices (Nakamura and Steinsson, 2008).

The results of these robustness tests do not alter the main findings we report in the paper and provide convergent evidence in support of H1 and H2 concerning the prevalence and rigidity of 0-ending prices in a setting where transaction convenience plays little role in affecting consumer behavior. See Appendix A for details of these tests.

### *6.2. Robustness tests with the US Dominick's scanner data*

To show that the prevalence and rigidity of 0-ending prices in the front-end-candies category is unique, we present evidence on the distribution and rigidity of 0-ending prices in each one of Dominick's 29 product categories. Then, focusing on the front-end-candies category, we show that our results are robust to the inclusion of outliers, to using an alternative definition of sale prices, to including observations from before 1991, and to the possibility that some of the price changes in the data could be an outcome of the way that prices in the dataset are calculated (Strulov-Shlain, 2022). We also show that we do not lose important information by bunching all non-0 and all non-9 ending prices together. See Appendix B for details of these robustness tests.

In addition, we conduct several robustness tests of our results on the demand for front-end-candies. These include using different weights for calculating the mean effects of the coefficients on price rigidity, including outliers in the regressions, including observations from before 1991 in the regressions, and removing the controls of quantity sold in store *s*, in week $t - 1$. The results are robust and provide additional support for H3 concerning the positive effect of 0-ending prices on the demand for conveniently purchased products (after controlling for the effect of prices' transaction convenience). See Appendix C for details of these robustness tests.

### *6.3. US Nielsen scanner data*

The Nielsen retail scanner dataset consists of information on weekly price, sales volume, and store merchandising reported by participating retail stores' point-of-sale systems in all US markets.[25] We use data on three product modules which contain products that are likely to be sold on the front-end display shelves (i.e., conveniently purchased products): chewing gums (8,190,505 observations on 181 products sold in 46,863 stores), bubble gums (2,827,275 observations on 10 products sold in 30,803 stores) and sugar-free chewing gums (51,571,803 observations on 733 products sold in 47,136 stores). We had to clean the dataset thoroughly (following the method detailed in Strulov-Shlain 2022), because the prices reported are quantity-adjusted weekly averages (and not transaction prices). We can therefore report only the results of estimating demand equations. The results are robust and provide additional support for H3 concerning the positive effect of 0-ending prices on the demand for conveniently purchased products using a different dataset. See Appendix D for details.

## 7. Conclusion

Multiple studies, including Jones (1896), Watkins (1911), Galbraith (1936), Deutsche Bundesbank (2002), Hoffman and Kurz-Kim (2006), Cornille and Stragier (2007), Glatzer and Rumler (2007), Levy and Young (2004, 2021), Young and Levy (2014), Knotek (2008, 2011), Bouhdaoui et al. (2014), Chen et al. (2019), Shy (2020), and Teber and Çevik (2022), find that 0-ending prices are common and rigid in settings where (1) consumers buy few items, (2) they pay in cash, and (3) there is high traffic. In such "convenience" locations, prices are set at 0-endings to minimize the number of coins needed to pay, and thus to facilitate transactions (i.e., *transaction convenience*).

We argue that in addition to transaction convenience, an additional factor, *cognitive convenience*, plays an important role in the popularity of 0-ending prices at convenience stores. Indeed, some existing studies suggest that the popularity of 0-ending prices may also be due to their *cognitive convenience*. However, these studies (Knotek 2011, 2008; Wieseke et al. 2016) do not distinguish cognitive convenience of 0-ending prices from their transaction convenience. Based on a literature review, we propose that 0-ending prices are more prevalent (H1), more rigid (H2), and increase demand more than other

[25] The data availability is subscription-based. Details about the data can be found at the Kilts Center of Marketing at the University of Chicago, at https://www.chicagobooth.edu/research/kilts/datasets/nielseniq-nielsen.

prices (H3) even after we control for the effect of prices' transaction convenience.

Using multiple datasets, we find strong empirical support for the three hypotheses. First, taking advantage of a unique legal setting in Israel, we show that 0-ending prices are common and rigid even when they have little effect on transaction convenience. Examining retail prices in Israel, we find that although 9-ending prices are common and rigid in Israeli supermarkets, the share of 0-ending prices at convenience stores in Israel exceeds 70 percent, greater than in the US (Knotek 2011). We also find that 0-ending prices are significantly more rigid than other prices in convenience stores.

We also show that 0-ending prices are common and rigid at a large U.S. supermarket chain in the front-end-candies category, for which cognitive convenience is likely to be important. We obtain these results after accounting for the popularity and the rigidity of 9-ending prices (Levy et al. 2011, Knotek 2019). In addition, we find that 0-ending prices have a positive effect on demand. Thus, our results suggest that retailers use 0-ending prices at convenience stores/settings because they increase sales.

We conduct many robustness checks for our results. In addition, we replicate the results on the positive effect of 0-ending prices on demand with the University of Chicago's AC Nielsen data for three products that are likely to be sold on the front-end display shelves. These robustness checks provide strong and convergent evidence for the prevalence, rigidity and demand-enhancing effect of 0-ending prices after controlling for prices' transaction convenience. Finally, our results rule out consumer antagonism (Blinder et al. 1998), and retailers' use of sub-optimal heuristics in pricing (DellaVigna and Gentzkow 2019, Huang et al. 2022) as possible explanations of our results.

Knotek (2008) notes that the rigidity of prices at convenience stores could be due to transaction or cognitive factors. While he favors the transaction convenience argument, he recognizes that his data does not allow him to distinguish between the two possibilities. Therefore, he calls for more "empirical work for other countries with distinct monetary systems" to study these issues (p. 25). Our work is a step in that direction, showing the popularity and the rigidity of 0-ending prices and their positive effect on demand at convenience stores in settings where they are used primarily for their cognitive convenience.

We thus demonstrate that 0-ending prices play an important role in settings where both consumers and producers value convenience. Using data from two countries, we show that 0-ending prices are common in such settings. This differs from the prevalence of 9-ending prices which are more common in large stores. We also find that 0-endng prices found in convenience settings remain rigid for extended periods of time, suggesting that monetary policy likely has a stronger effect on sectors of the economy where convenient goods and services are traded. Finally, we show that 0-ending prices are correlated with greater demand, suggesting that such prices are optimal from the viewpoint of sellers: by setting 0-ending prices in convenience settings, retailers sell more.

Our work further suggests that although retailers often use prices to obfuscate information (Gabaix and Laibson 2006, Chen et al. 2008, Chakraborty et al. 2015, McShane et al. 2016, Levy et al. 2020, Snir and Levy 2021), in some settings it is advantageous to facilitate information processing. In settings where shoppers make spontaneous decisions on a small number of goods, reducing the cognitive effort that is needed to make decisions can increase the likelihood of a sale. In such settings, price information transparency can benefit everyone. Thus, 0-ending prices might remain popular, and their effect on price rigidity might persist even in settings where transactions are processed electronically (Chenavaz et al. 2018, Gorodnichenko and Talavera 2017).

Because of data limitations, our scanner price data analysis was limited to products in the front-end candy category. However, in addition to candy products, most large US retailers sell dozens of other convenience products through front-end point-of-purchase display shelves. These include small tools, batteries, books and magazines, shaving razor blades, snacks, individual-sized bottled drinks and sodas. Future studies should therefore focus on assessing the prevalence, the rigidity, and the demand effects of 0-ending prices for these products as well.

Figure 1. The distribution of price endings in Israeli superstores and convenience stores

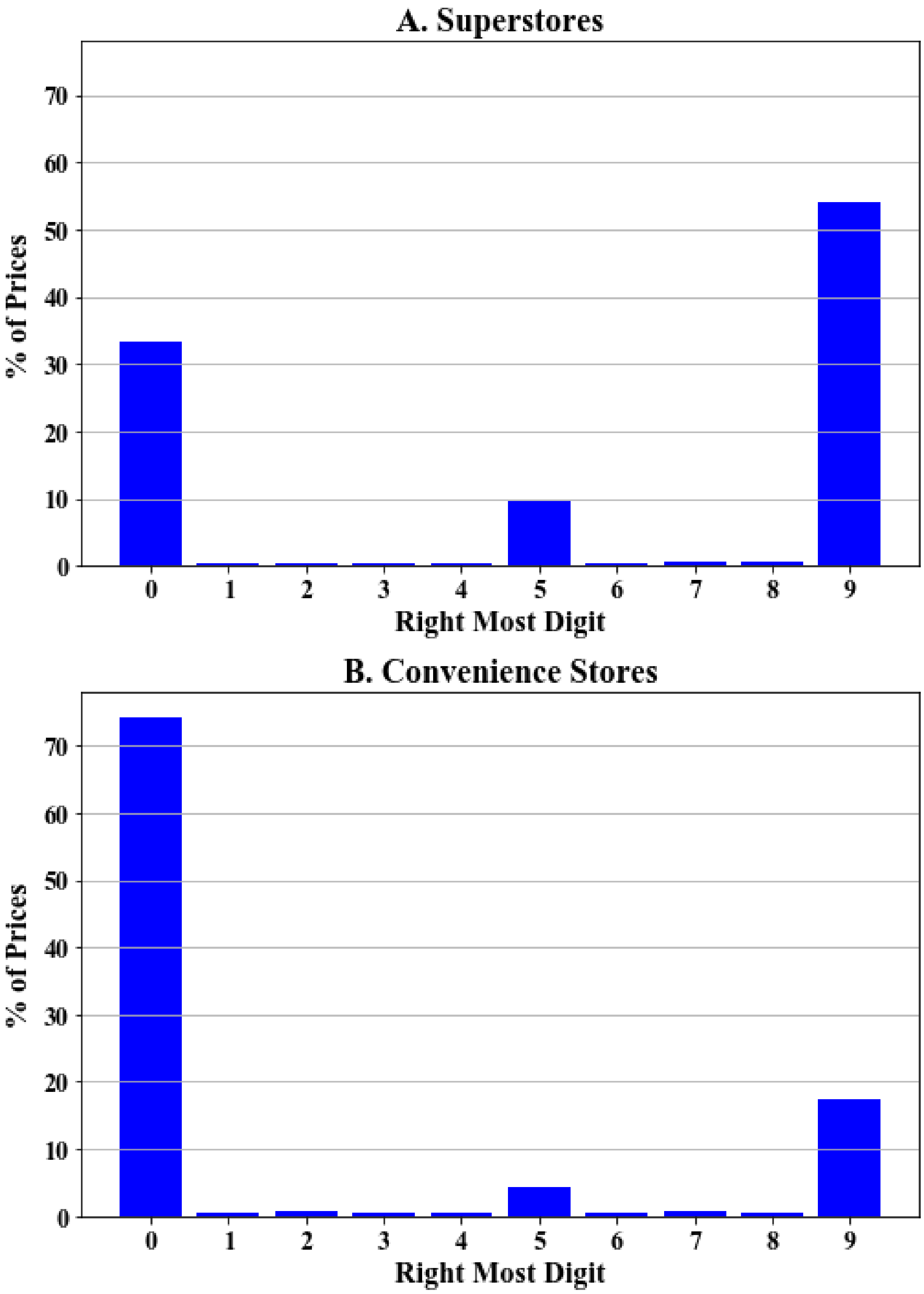

Note: The figures are based on the data of Israel's Central Bureau of Statistics for 125 ELIs of consumer goods, covering the period January 1999–August 2014.

Figure 2. Shares of 0-ending and 9-ending prices in Israeli superstores and convenience stores

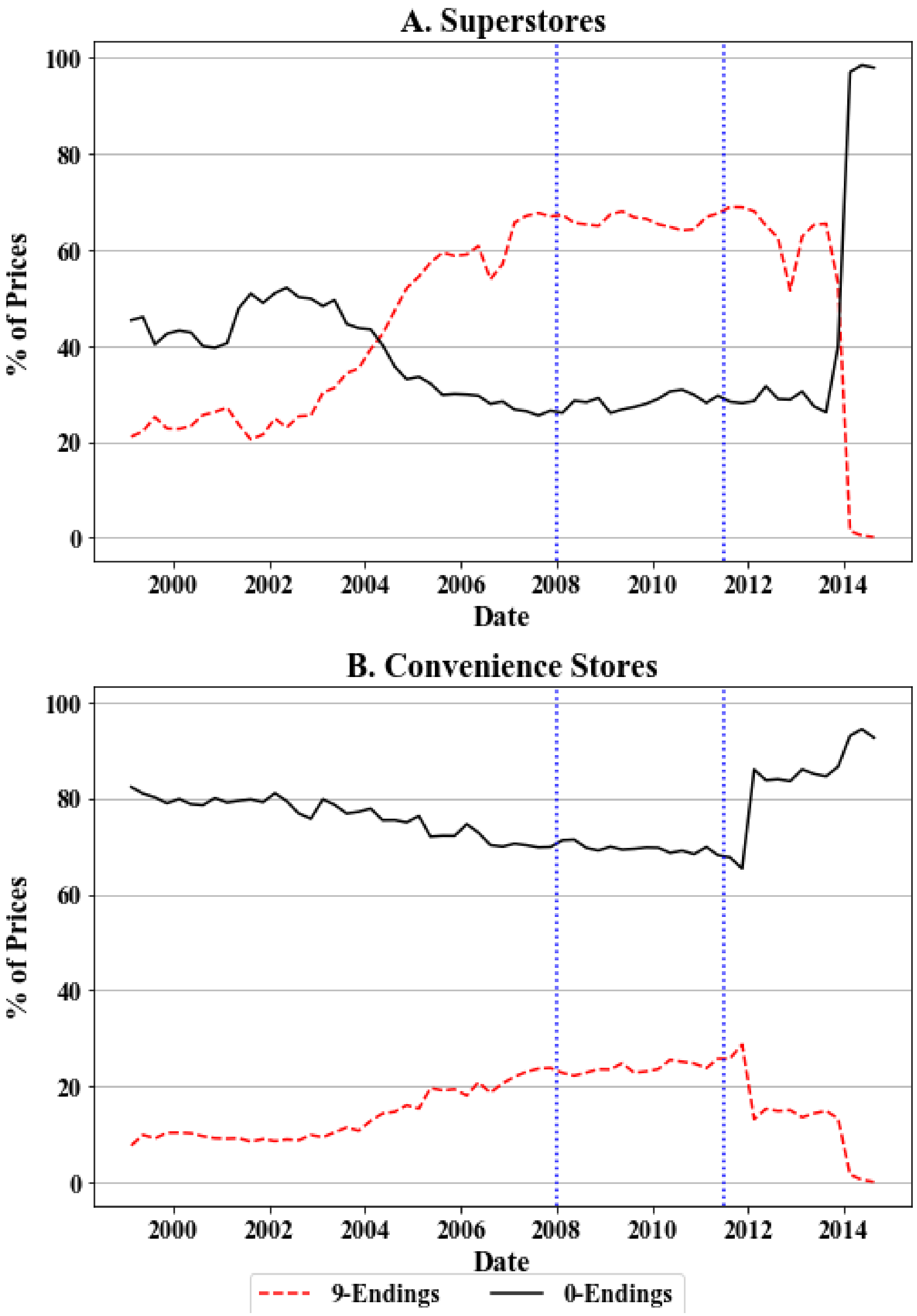

Notes: The figures are based on the data of Israel's Central Bureau of Statistics for 125 ELIs of consumer goods, covering the period January 1999–August 2014. The first vertical line indicates the date on which 5-agora coins ceased to be legal tender (i.e., January 1, 2008). The second vertical line indicates the beginning of the "cottage cheese protests" (i.e., June 15, 2011).

Figure 3. The likelihood of a price change at convenience stores and superstores, Israel

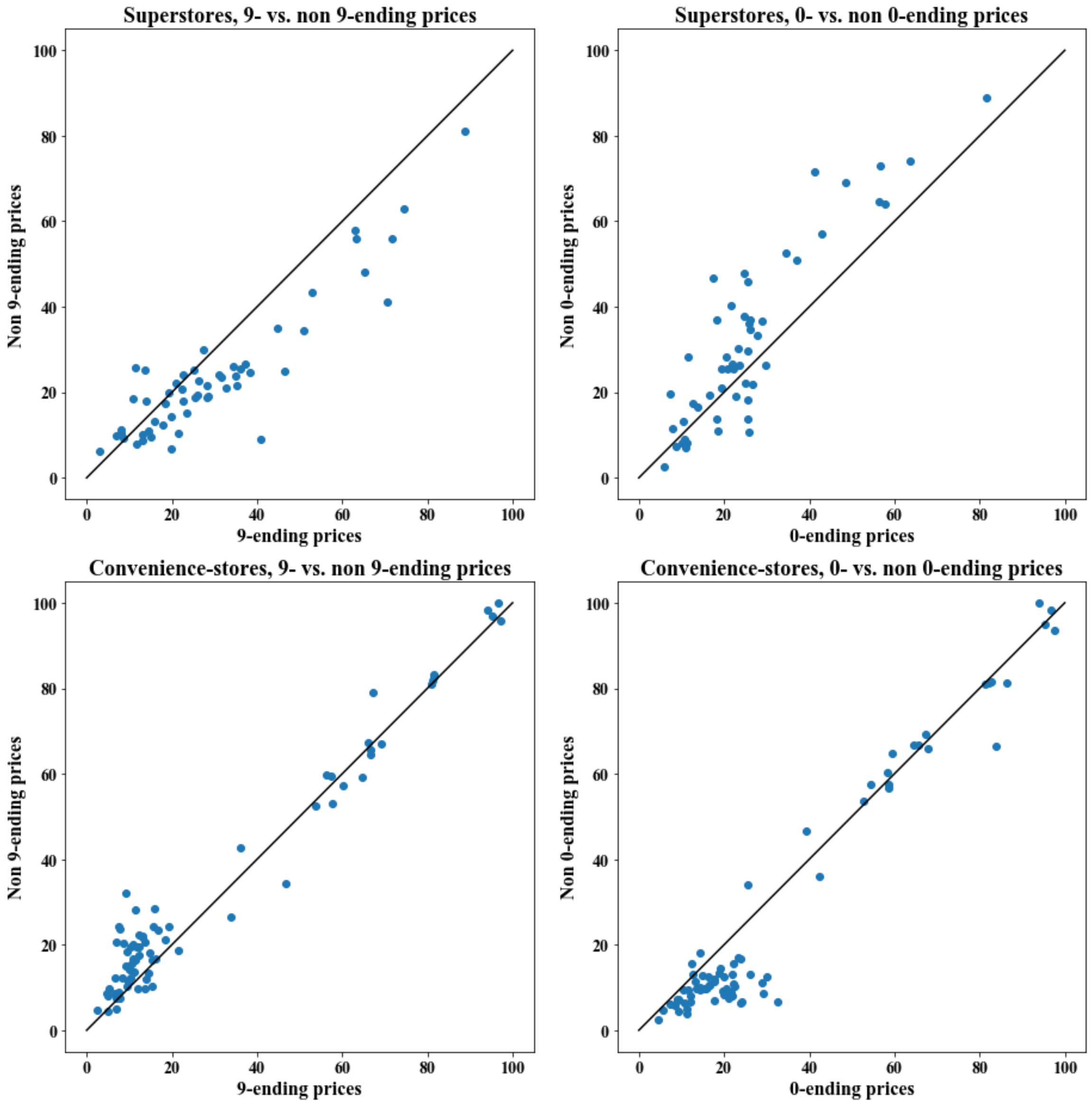

Notes: In each panel, the *x*-axis gives the average percentage of the 9-ending or 0-ending prices that change in a given week. The *y*-axis gives the average percentage of non 0-ending or non 9-ending prices that change in a given week. On each panel, there are 125 dots representing 125 product categories. Dots lying below (above) the diagonal lines indicate that the *x*-variable is more (less) rigid, i.e., it is less (more) likely to change, than the *y*-variable in the corresponding category.

Figure 4. Price ending distribution: Front-end-candies category vs. the remaining 28 categories, Dominick's, U.S.

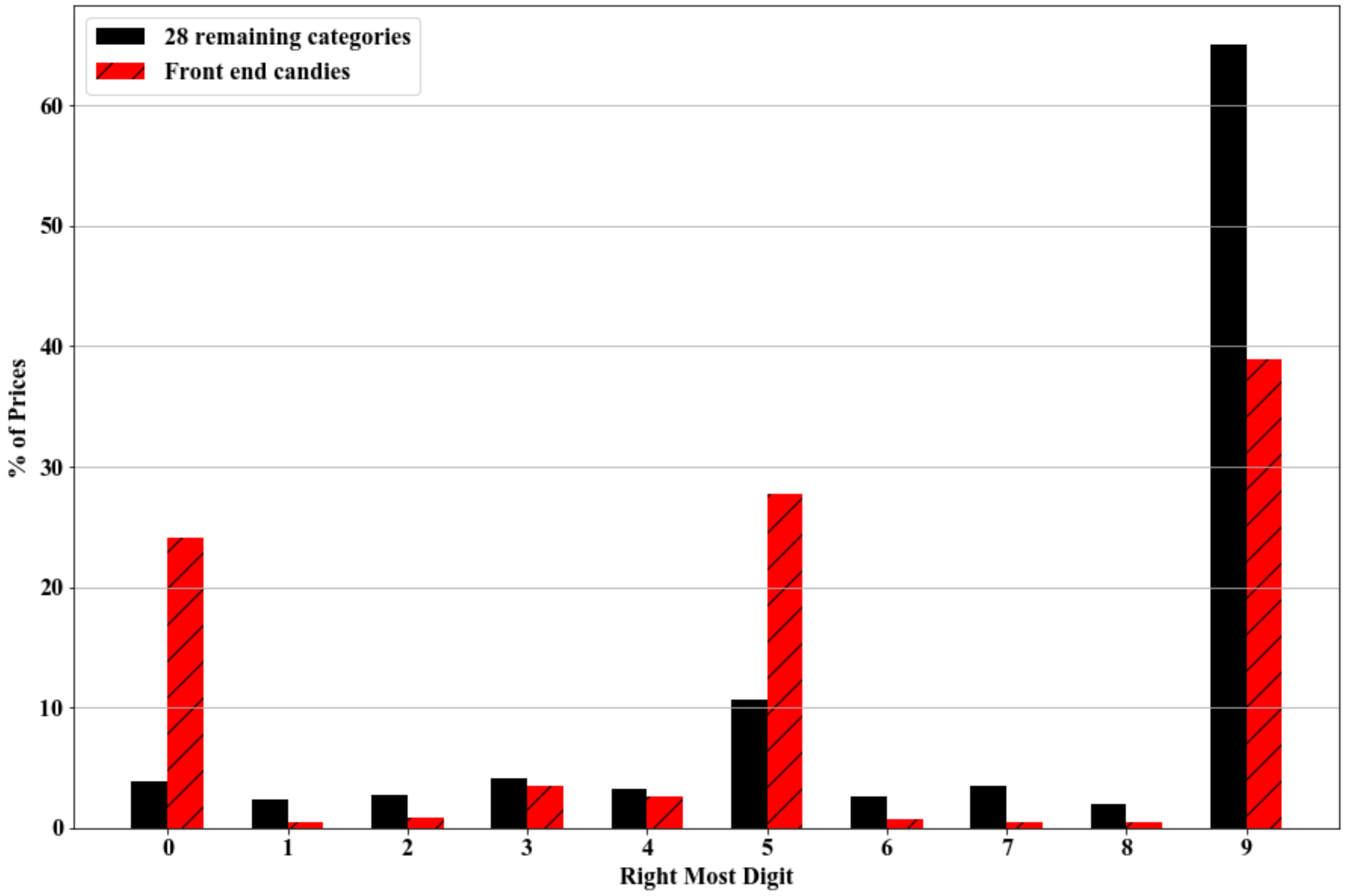

Figure 5. Likelihood of a price change by price ending: Front-end-candies category, Dominick's, U.S.

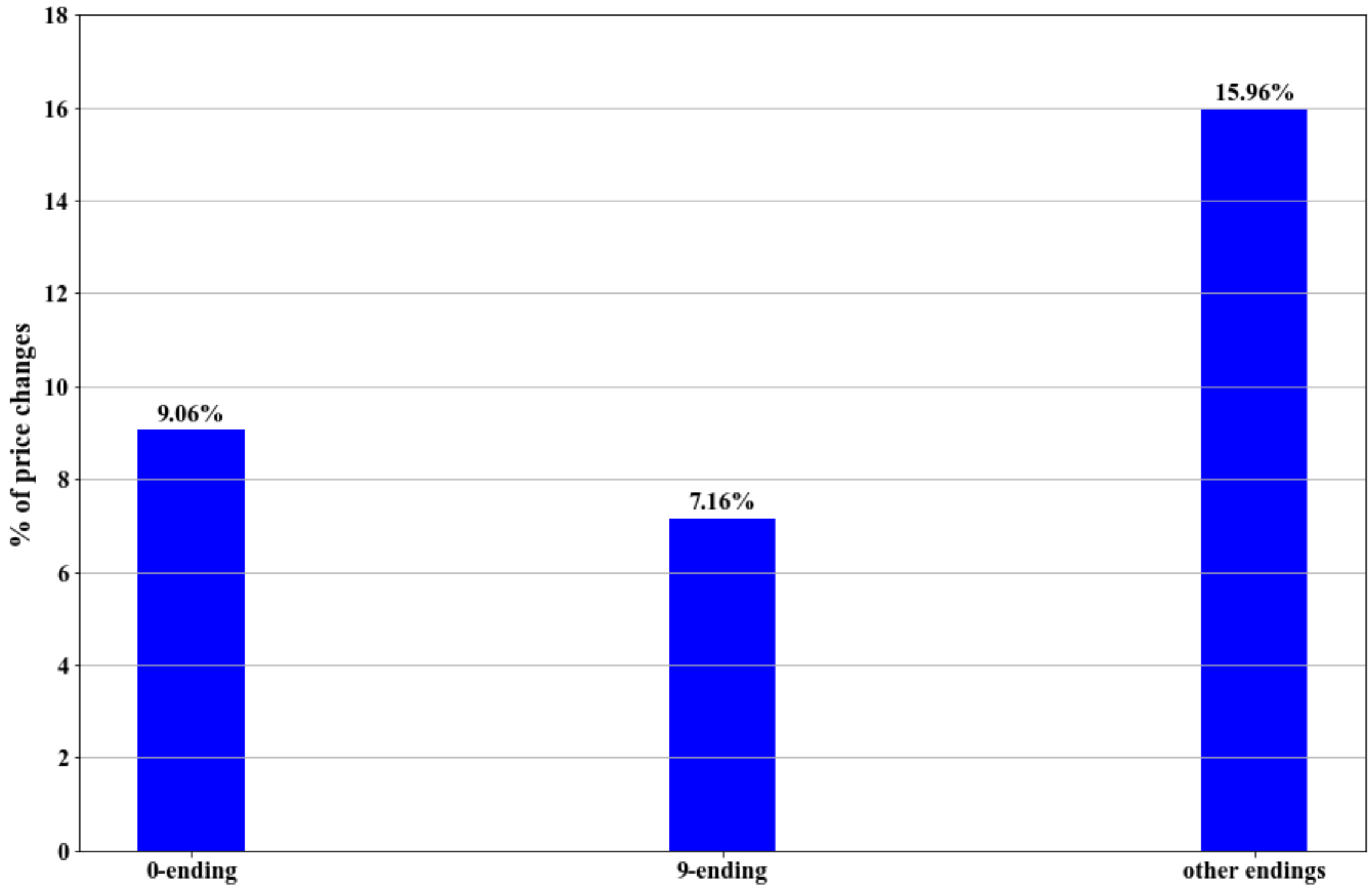

Figure 6. Demand for Breath Savers Wintergreen Sugar-Free Mints, Store No. 101, Dominick's

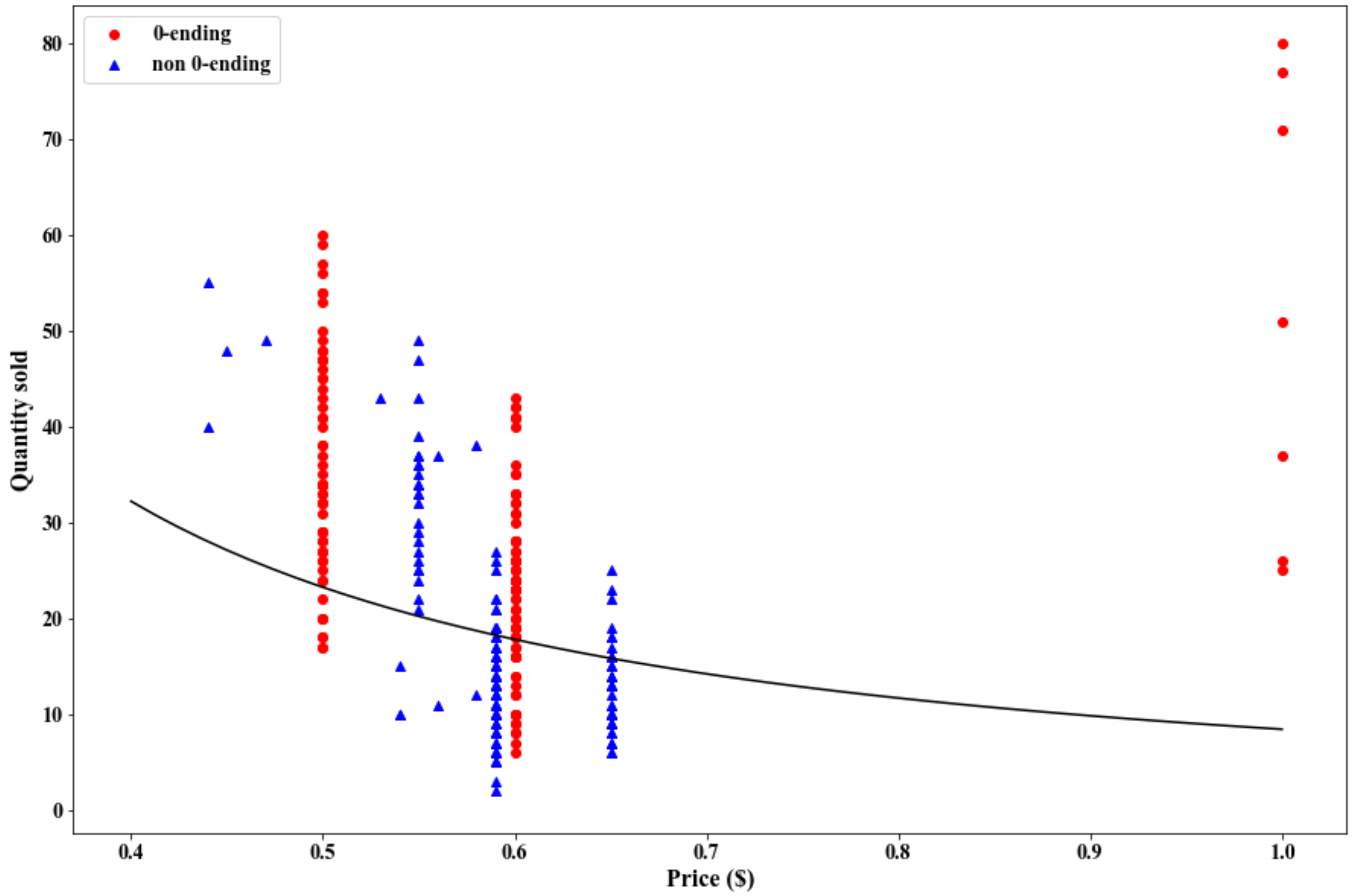

Notes: Weekly units of Breath Savers Wintergreen Sugar-Free Mints (UPC=1900000093) at Store No. 101, Dominick's, U.S. 0-ending prices are marked with red circles. All other ending prices are marked with blue triangles.

**Table 1**
Regressions of the probability of a price change: Convenience stores, Israel

| | Baseline Regression (1) | Convenience Store (2) | Added Controls (3) | Sale Prices (4) | Maximum *P* = NIS10 (5) | All Other Prices (6) | Consumer Antagonism (7) | Cottage Cheese Protests (8) |
|---|---|---|---|---|---|---|---|---|
| 0-ending | –0.072*** (0.009) | –0.003 (0.014) | –0.027** (0.013) | –0.025* (0.013) | 0.073* (0.041) | –0.093*** (0.016) | –0.037*** (0.013) | –0.038*** (0.013) |
| 9-ending | –0.036*** (0.09) | –0.070*** (0.000) | –0.085*** (0.012) | –0.084*** (0.012) | –0.067** (0.022) | –0.120*** (0.000) | –0.054*** (0.012) | –0.055*** (0.015) |
| Transaction-inconvenience | 0.006*** (0.001) | 0.002** (0.001) | 0.009*** (0.001) | 0.009*** (0.001) | 0.024*** (0.006) | 0.005*** (0.001) | 0.011*** (0.001) | 0.011*** (0.001) |
| Convenience store × 0-ending | | –0.057*** (0.013) | –0.052*** (0.013) | –0.052*** (0.013) | –0.122*** (0.039) | –0.054*** (0.014) | –0.045*** (0.012) | –0.044*** (0.013) |
| Convenience store × 9-ending | | 0.045*** (0.011) | 0.045*** (0.011) | 0.044*** (0.011) | 0.041** (0.021) | 0.020 (0.013) | 0.033*** (0.012) | 0.033*** (0.012) |
| Convenience store × transaction-inconvenience | | 0.003** (0.001) | 0.002* (0.0001) | 0.002* (0.0001) | –0.005 (0.006) | 0.000 (0.001) | 0.001 (0.001) | 0.001 (0.001) |
| 5-ending | | | –0.035*** (0.011) | –0.033*** (0.011) | –0.022** (0.011) | –0.084** (0.018) | –0.035*** (0.011) | –0.035*** (0.011) |
| Sale dummy | | | | 0.441*** (0.009)_ | 0.387*** (0.010) | 0.511*** (0.014) | 0.441*** (0.009)_ | 0.440*** (0.009)_ |
| Post-2008 dummy × 0-ending | | | | | | | 0.033* (0.018) | 0.035* (0.020) |
| Post-2008 dummy × 9-ending | | | | | | | –0.046*** (0.010) | –0.047*** (0.011) |
| Post-2008 dummy × Transaction-inconvenience | | | | | | | –0.004*** (0.001) | –0.004*** (0.001) |
| Post-2008 dummy × convenience × 0-ending | | | | | | | –0.019 (0.019) | –0.024 (0.022) |
| Post-2008 dummy × convenience × 9-ending | | | | | | | 0.012 (0.010) | 0.012 (0.010) |
| Post-2008 dummy × convenience × transaction-inconvenience | | | | | | | 0.001** (0.001) | 0.001 (0.001) |
| Cottage cheese protest dummy × 0-ending | | | | | | | | –0.001 (0.019) |
| Cottage cheese protest dummy × 9-ending | | | | | | | | 0.009 (0.017) |
| Cottage cheese protest dummy × transaction-inconvenience | | | | | | | | 0.001 (0.001) |
| Cottage cheese protest dummy × convenience × 0-ending | | | | | | | | 0.020 (0.016) |
| Cottage cheese protest dummy × convenience × 9-ending | | | | | | | | –0.001 (0.012) |
| Cottage cheese protest dummy × convenience × Transaction-inconvenience | | | | | | | | –0.003** (0.001) |
| Constant | 0.259***<br>(0.002) | 0.298***<br>(0.002) | 0.286***<br>(0.073) | 0.284***<br>(0.073) | 0.010<br>(0.090) | 0.412***<br>(0.074) | 0.283***<br>(0.005) | 0.285***<br>(0.005) |
| $R^2$ | 0.299 | 0.303 | 0.303 | 0.306 | 0.313 | 0.275 | 0.307 | 0.307 |
| Number of observations | 564,742 | 564,742 | 564,742 | 564,742 | 253,984 | 310,758 | 564,742 | 564,742 |

Notes

1. The table reports the estimation results of a set of linear probability models.
2. The dependent variable is a dummy that equals 1 if the price has changed in a given month.
3. The independent variables are: 0-ending – a dummy which equals 1 if the previous price was 0-ending, 9-ending – a dummy which equals 1 if the previous price was 9-ending, transaction-inconvenience – the minimum number of coins necessary for paying a price in cash, convenience store dummy – an outlet that is defined as a convenience store, 5-ending – a dummy which equals 1 if the previous price was 5-ending, sale dummy – a dummy for sale prices in month $t$ – 1, post-2008 dummy – a dummy that equal 1 for observations after January 2008, and "cottage cheese protest" dummy – a dummy that equal 1 for observations after August 2011.
4. All the regressions also include controls for product categories, for the 6 regions of Israel and for the quarter to which the observation pertains.
5. Columns 2–8 also include the dummy for convenience stores.
6. Columns 3–8 also include: price level – the price in month $t$ – 1 rounded to the nearest NIS, population – the number of inhabitants in the town where the store is located (in logs), socio-economic score/rank – calculated by the Israeli Central Bureau of Statistics for the town where the store is located, share of women – the share of women in the town where the store is located, distance from Tel-Aviv – the distance of the town where the store is located from Tel-Aviv (logs, in km), and share of minority groups – the share of minority groups in the town where the store is located.
7. Columns 7–8 also include the post-2008 dummy and column 8 also includes the "cottage cheese protest" dummy.
8. Standard errors are clustered at the level of the store where the observation was sampled.
9. * $p < 10\%$, ** $p < 5\%$, *** $p < 1\%$

**Table 2**

Regressions of the probability of a price change in the front-end-candies category: Dominick's, U.S.

| | All Observations (1) | All Observations (2) | Regular Prices (3) | Sale Prices (4) |
|---|---|---|---|---|
| 0-ending price | –0.021*** (0.002) | –0.104*** (0.003) | –0.067*** (0.003) | –0.109*** (0.004) |
| 9-ending price | –0.137*** (0.002) | –0.156*** (0.003) | –0.118*** (0.003) | 0.046*** (0.005) |
| Transaction-inconvenience score | 0.045*** (0.001) | 0.022*** (0.001) | 0.021*** (0.001) | –0.052*** (0.001) |
| Quarter-multiple | | 0.004** (0.002) | –0.005*** (0.001) | –0.018*** (0.004) |
| Price level | | –0.051*** (0.002) | –0.030*** (0.001) | 0.021*** (0.004) |
| Absolute value of the % change in wholesale price | | 0.011*** (0.000) | 0.009*** (0.000) | 0.010*** (0.000) |
| 5-ending price | | –0.141*** (0.002) | –0.101 (0.003) | 0.081*** (0.005) |
| Sale price | | 0.438*** (0.002) | | |
| Constant | 0.073*** (0.002) | 0.017*** (0.002) | 0.099*** (0.003) | 0.650*** (0.005) |
| $R^2$ | 0.038 | 0.333 | 0.181 | 0.098 |
| No. of observations | 3,708,902 | 3,686,663 | 3,214,357 | 472,306 |

Notes

1. The table presents the results of a linear probability model for the probability of a price change.
2. The dependent variable is a dummy, which equals 1 if the price of good *i* in store *j* changed in week *t*.
3. The independent variables are: 0-ending – a dummy which equals 1 if the previous price was 0-ending, 9-ending – a dummy which equals 1 if the previous price was 9-ending, quarter-multiple – a dummy which equals 1 if the previous price was either 25- or 75-ending, transaction-inconvenience score – the minimum number of coins needed to pay the previous price, price level – the previous price rounded to the nearest dollar, sale price – a dummy for the previous price being a sale price, absolute value of the percentage change in the wholesale price – the percentage absolute change in the wholesale price, and 5-ending price – a dummy which equals 1 if the previous price was 5-ending.
4. The regressions in columns (1)–(2) use all observations, the regression in column (3) uses only observations on regular prices, and the regression in column (4) uses only observations on sale prices.
5. The regressions also include sub-category×stores×week fixed effects.
6. In columns (2)–(4), we drop 22,239 observations (about 0.6% of the total) that have an absolute value of the percentage change in the wholesale price above 150%.
7. The $R^2$ row reports the overall $R^2$ of the fixed effects regression.
8. Robust standard errors, clustered at the store-product level, are reported in parentheses.
9. * $p$ < 10%, ** $p$ < 5%, *** $p$ < 1%

**Table 3**

Regressions of the effects of price endings on demand in the front-end-candies category, Dominick's, U.S.

| | Average | Quantity adjusted average | No. of positive coefficients | No. of Negative coefficients | No. of positive and significant coefficients | No. of negative and significant coefficients | No. of observations |
|---|---|---|---|---|---|---|---|
| **Panel A** | | | | | | | |
| 0-ending price | 0.173 | 0.206 | 51 | 29 | 43 | 13 | 1,988,759 |
| 9-ending price | –0.009 | –0.026 | 35 | 45 | 28 | 38 | |
| Transaction-inconvenience score | 0.067 | 0.074 | 64 | 16 | 53 | 10 | |
| Log(price) | –0.264 | –0.299 | 19 | 61 | 13 | 56 | |
| **Panel B** | | | | | | | |
| 0-ending price | 0.122 | 0.136 | 44 | 36 | 28 | 17 | 1,988,759 |
| 9-ending price | –0.012 | –0.015 | 34 | 43 | 16 | 30 | |
| Transaction-inconvenience score | 0.064 | 0.083 | 58 | 22 | 43 | 14 | |
| Log(price) | –0.042 | –0.046 | 41 | 39 | 33 | 32 | |
| **Panel C** | | | | | | | |
| 0-ending price | 0.145 | 0.167 | 49 | 31 | 27 | 18 | 1,847,672 |
| 9-ending price | 0.026 | 0.016 | 48 | 29 | 34 | 16 | |
| Transaction-inconvenience score | 0.043 | 0.058 | 61 | 19 | 46 | 11 | |
| Log(price) | 0.031 | 0.006 | 49 | 31 | 40 | 20 | |
| **Panel D** | | | | | | | |
| 0-ending price | 0.006 | 0.202 | 33 | 16 | 17 | 9 | 140,606 |
| 9-ending price | –0.029 | –0.088 | 26 | 23 | 9 | 9 | |
| Transaction-inconvenience score | 0.038 | 0.149 | 33 | 16 | 19 | 7 | |
| Log(price) | –0.244 | –0.464 | 3 | 46 | 0 | 39 | |

Notes

1. The table summarizes the estimation results of reduced form, product-level demand equations for 80 products in the Dominick's front-end-candies' category, a total of 80 regressions.
2. The figures in the table are the average quantity-adjusted coefficients computed over the 80 regression equations that we estimate.
3. In each regression, the dependent variable is the log of the quantity of product *q* sold at store *s*, in week *t*.
4. The independent variables include a dummy for 0-ending price (1 if the previous price ends with 0, 0 otherwise), a dummy for 9-ending prices (1 if the previous price ends with 9, 0 otherwise), the transaction-inconvenience score (the minimum number of coins needed to pay the previous price), the log of the price, the log of the average price of other products in the same sub-category, the quantity sold in the same store in the previous week, and fixed effects for years and months, for stores, and for holidays, which include Christmas, New Year, Presidents Day, Easter, Memorial Day, 4th of July, Labor Day, Halloween, and Thanksgiving.
5. The log of the wholesale price is used as the instrument for the price.
6. In panels A–B, we report the estimation results using all observations.
7. In panel B, the regression equations also include a dummy for sale prices, which we identify using a sales filter, a dummy for 5-ending prices (1 if the previous price ends with 5, 0 otherwise), and a quarter-multiples (l if the previous price ends with 25 or 75, 0 otherwise).
8. In panel C, we use observations on regular prices only, while in panel D, we use observations on sale prices only, where we identify sale prices using a sale filter.
9. We use only observations on products for which we have at least 1,000 observations on sale prices, leaving us with 49 products.

**Online Supplementary Web Appendix**

# Zero-Ending Prices, Cognitive Convenience, and Price Rigidity

Avichai Snir
Department of Economics
Bar-Ilan University
Ramat-Gan 52900, Israel
avichai.snir@gmail.com

Haipeng (Allan) Chen
Department of Marketing and Supply Chain
University of Kentucky
Lexington, KY 40506, USA
allanchen@uky.edu

Daniel Levy
Department of Economics
Bar-Ilan University
Ramat-Gan 52900, Israel,
Department of Economics
Emory University
Atlanta, GA 30322, USA,
RCEA, Italy,
ICEA, Canada, and
ISET at TSU, Georgia
Daniel.Levy@biu.ac.il

October 2, 2022

**Table of Contents**

## Appendix A. Robustness tests for the Israeli CPI data

In the paper, we conduct several tests showing that 0-ending prices are more rigid than other prices at Israeli convenience stores.[1] Below we provide the results of several additional robustness tests, to show that the results we report in the paper are not driven by outliers.

In Table A1, we estimate regressions similar to the ones we report in column 4 of Table 1 in the paper. These regressions include a full set of control variables, in addition to the main independent variables. The dependent variable is a dummy that equals 1 if the price has changed and 0 otherwise. The list of independent variables includes: a 0-ending price dummy which equals 1 if the previous price ended in 0, a 9-ending price dummy which equals 1 if the previous price ended in 9, a transaction-inconvenience index which equals the minimum number of coins/notes needed to pay the price, a dummy for convenience stores, an interaction of the convenience store dummy with 0-ending dummy, with 9-ending dummy, and with the transaction-inconvenience index, the price level which is calculated as the previous period price rounded to the nearest NIS, the log of the population of the town where the store is located, the socio-economic score of the town where the store is located, the share of women in the town where the store is located, the log of the distance of the town where the store is located from Tel-Aviv, the share of minority groups in the town where the store is located, and a sale dummy which equals 1 if the previous period price was a sale price. The regression also controls for product categories, the six regions of Israel, and the quarter in which the observation was taken. Standard errors are clustered at the level of the store where the observation was taken.

In column 1, we restrict the sample to goods with prices ≤ NIS 200. It is likely that higher prices are usually not paid with cash (Chen et al. 2019, Shy 2020), as the highest denomination bill in Israel is NIS 200. We find that the coefficient of the 0-ending price dummy is positive and significant ($\beta = 0.010, p < 0.01$), while the coefficient of the 9-ending prices dummy is negative and significant ($\beta = -0.045, p < 0.01$). We therefore find that when we restrict the sample to prices up to NIS 200, then at superstores 0-ending prices are not more rigid than other prices, but 9-ending prices are.

[1] Following Knotek (2011), we classify a store as a *superstore* if the CBS flags it as a supermarket, chain store, department store, or a drugstore. We classify a store as a *convenience store* if the CBS flags it as a small grocery store, gas station, kiosk, convenience shop, or a specialty shop such as a bakery, fruits/vegetables store, etc.

The coefficient of the transaction inconvenience index is positive and significant ($\beta = 0.017, p < 0.01$), suggesting that an increase in the number of coins needed to pay the price is associated with a greater likelihood of a price change.

The interaction of the 0-ending dummy with the convenience-stores dummy is negative and significant ($\beta = -0.063, p < 0.01$). In convenience stores therefore, 0-ending prices are more rigid than other prices even when we restrict the sample to prices of up to NIS 200 ($F = 92.9, p < 0.00$). The interaction of the 9-ending dummy with the convenience-stores dummy is positive and significant ($\beta = 0.030, p < 0.01$), implying that although 9-ending prices are also more rigid than other prices in convenience stores ($F = 3.28, p < 0.08$), they are less rigid in than in superstores. Further, in convenience stores, 9-ending prices are less rigid than 0-ending prices ($F = 35.1, p < 0.01$).

In column 2, we further restrict the sample. This time, we restrict the sample to prices that are ≤ NIS 4, to match it with the price level of the products we study at Dominick's ($1).[2] We find that in superstores, 0-ending prices are not more rigid than other prices ($\beta = 0.089, p < 0.01$), while 9-ending prices are ($\beta = -0.038, p < 0.01$). An increase in the number of coins needed to pay a price is associated with a higher likelihood of a price change ($\beta = 0.022, p < 0.01$), i.e., lower price rigidity.

The coefficient of the interaction of the 0-ending dummy with the convenience stores dummy is negative and significant ($\beta = -0.113, p < 0.01$), implying that in convenience stores 0-ending prices are more rigid than other prices ($F = 3.33, p < 0.07$). The coefficient of the interaction of the 9-ending dummy with the convenience stores dummy is positive ($\beta = 0.003, p < 0.01$). After taking into account the main effect, we still find that for prices smaller than NIS 4, 9-ending prices are more rigid than other prices also in convenience stores ($F = 5.68, p < 0.02$). For this level of prices, the rigidity of 0- and 9-ending prices is not statistically different ($F = 1.09, p > 0.29$).

[2] During the sample period, the average NIS/$ exchange rate was NIS 4.09 per $1.

In column 3, we restrict the sample to prices that can be paid with a single coin or a single note, e.g., all prices in the range of 4.95–5.04, or 19.95–20.04.[3] These are the most convenient prices and, therefore, the ones that are most likely to be paid with cash (Knotek 2008, 2011, 2019; Chen et al. 2019, Shy 2020). We find that in superstores, 0-ending prices are not more rigid than other ending prices ($\beta = 0.002, p < 0.01$), while 9-ending prices are ($\beta = -0.054, p < 0.01$). The interaction of convenience stores and 0-ending prices is negative and significant ($\beta = -0.071, p < 0.01$), while the interaction with 9-ending prices is positive and significant ($\beta = 0.018, p < 0.01$). Thus, when we restrict the sample to prices that can be paid with one coin/note, we find that in convenience stores both 0-ending prices ($F = 12.3, p < 0.01$) and 9-ending prices ($F = 3.4, p < 0.07$) are more rigid than other prices. 0-ending prices, however, are more rigid than 9-ending prices ($F = 12.0, p < 0.01$).

In column 4, we restrict the sample to regular prices, by removing the sale and bounce-back prices. A large stream of literature in macroeconomics has shown that at the aggregate level, regular price changes are more important than temporary price changes (i.e., sales). See, for example, Nakamura and Steinsson (2008), Eichenbaum et al. (2011), and Midrigan (2011). In this regression, we therefore check if our results remain qualitatively unchanged if we exclude the sale prices.

The results are again similar to the ones we report above. When we restrict the sample to regular prices, we find that 0-ending prices are not more rigid than other prices in superstores ($\beta = 0.008, p < 0.00$). The interaction of 0-ending prices with convenience-stores, however, is negative and significant ($\beta = -0.053, p < 0.01$), implying that in convenience-stores, 0-ending prices are more rigid than other prices ($F = 46.8, p < 0.01$). 9-ending prices are more rigid than other prices in superstores ($\beta = -0.047, p < 0.01$). The interaction of 9-ending prices with convenience-stores, however, is positive and significant ($\beta = 0.036, p < 0.01$), implying that in convenience-stores, 9-ending prices are not more rigid than other prices ($F = 2.5, p < 0.12$).

[3] This range of prices refers to the period after 2008. Before 2008, Prices in the range 4.98–5.02 could have been paid with a single coin.

In column 5, we estimate the regression using the full set of data, but replacing the linear probability model with a probit model. We find that using a non-linear estimation technique does not affect the results qualitatively. 0-ending prices are not more rigid than other prices in superstores, but they are more rigid than other prices in convenience stores. 9-ending prices are more rigid than other prices in superstores and in convenience stores. In convenience stores, however, 0-ending prices are more rigid than 9-ending prices ($\chi^2 = 37.2, p < 0.01$).

In column 6, we focus on convenience stores and we add dummy variables for each possible price ending. We find that the coefficient of 0-ending is negative and statistically significant ($\beta = -0.056, p < 0.01$). Furthermore, in column 1 of Table A2, we compare the coefficient of 0-ending prices with the coefficients of prices with other endings. We find that 0-ending prices are significantly more rigid than prices with any other ending.

For completeness, in column 7 we focus on superstores. We find that the coefficient of 0-ending prices is negative and statistically significant ($\beta = -0.044, p < 0.01$). However, when we compare the coefficient of 0-ending prices with the coefficients of prices with other endings (column 2 of Table A2), we find that in most cases, the difference is not statistically significant. In other words, in convenience stores, prices that end in 0 are less likely to change than prices that end with any other ending. In superstores, 0-ending prices are not particularly rigid.

In column 8 we check if the results are robust to changing the definition of convenience stores. In the paper, we define a store as a convenience store if the CBS flags it as a convenience store, a small grocery, a kiosk, an open market stall, or a specialty store. For this test, we define a store as a convenience store only if the CBS defines it as a convenience store. We then estimate a regression using data only on observations from convenience stores. We had to drop the dummy for sales because the number of sales in convenience stores is too low. We find that both the coefficient of 0-ending prices ($\beta = -0.132, p < 0.01$) and of 9-ending prices ($\beta = -0.141, p < 0.01$) are negative and statistically significant. The difference between the two coefficients are not statistically different ($F = 0.27, p > 0.60$).

An alternative explanation for the frequent use of 0-ending prices at convenience stores is that retailers might be using them to save compuatation time. Before 2008, if a consumer bought three products that have a 9-ending price and pay in cash, then the cashier would most likely have to give him a 5-agora coin as a change. For example, if the consumer bought three products costing NIS 4.99, then the total price would be NIS 14.97 which would be rounded down to NIS 14.95. Therefore, if the consumer was to pay NIS 15 in cash, the retailer would have to give him a 5-agora coin as change. This implies that using 0-ending prices could simplify the computations and enhance the transaction speed also at convenience stores.

To check the relevance of this type of transaction convenience, we focus on the sample of convenience stores, and conduct two tests. First, we divide the convenience stores into those that are likely to have a till—convenience stores, small groceries and specialty stores, and those that are unlikely to have a till—kiosks and open market stalls. If transaction convenience played a significant role in the prevalence of 0-ending prices, it is likely that stores that do not have a till would be more likey to use 0-ending prices than stores that have a till, since the till simplifies the calculations that the cashier has to perform.

We find that 90.0% of the prices at stores that are unlikely to have a till are 0-ending, compared to 63.8% at convenience stores that are likely to have a till. The difference is statistically significant (Wilcoxon rank-sum test $z = 195.1, p < 0.01$). Thus, it seems that transaction convenience might have played some role in the setting of 0-ending prices. Nevertheless, convenience stores that do not have a till still had over 60% 0-ending prices, suggesting that transaction convenience was not the only consideration for using 0-ending.

We can also test whether having a till affected the price rigidity of 0-ending prices. We focus on the sample of convenience stores, and estimate a regression with the full set of control variables, as above. The main independent variables are 0-ending dummy, 9-ending dummy, a dummy for stores without a till, and interactions of the dummy for stores without a till with the dummies for 0-ending and for 9-ending prices. The estimation results are given in the first column of Table A3.

We find that both 0-ending prices ($\beta = -0.062, p < 0.01$) and 9-ending prices ($\beta = -0.035, p < 0.01$) are more rigid than other prices at convenience stores. However, the coefficient of the interaction of the 0-ending dummy with the dummy for stores without a till is not statistically significant ($\beta = -0.026, p > 0.74$). The interaction of the 9-ending dummy with the dummy for stores without a till is also not statistically significant ($\beta = -0.070, p > 0.39$). The interaction of the transaction inconvenience score with the dummy for stores without a till is positive and statistically significant ($\beta = 0.010, p < 0.01$). Therefore, it seems that stores that do not use a till maintain transaction-convenient prices unchanged for longer periods than convenience stores with a till. The absence of a till does not affect, however, the duration of 0-ending and 9-ending prices, suggesting that the main reason for setting such prices is not transaction convenience.

As a further test, we look at the shares and rigidity of 0-ending prices before and after 2008. After 2008, the 5-agora coin ceased to be a legal tender. Consequently, a consumer had to buy more 9-ending products before the retailer had to give him change: Before 2008, a consumer had to buy three 9-ending products before the price was rounded down to NIS 0.95. After 2008, the consumer had to buy 6 products before the price was rounded down to NIS 0.90. Thus, the transaction inconvenience of non-round prices became even less acute after 2008. Therefore, if the main reason for using 0-ending prices was transaction inconvenience, then after 2008 the share of 0-ending prices should have decreased significantly.

We find that the share of 0-ending prices indeed decreased slightly. At convenience stores with a till, it decreased from 65.5% to 61.8% (Wilcoxon rank-sum test $z = 22.8, p < 0.01$). At convenience stores without a till, it decreased from 91.7% to 87.2% (Wilcoxon rank-sum test $z = 29.4, p < 0.01$).

Thus, when 0-ending became less important for transaction convenience, their share has decreased, but it stayed high at stores both with and without tills. It is another indication that although transaction convenience played some role in the high share of 0-ending prices, it is not the only reason.

In column 2 of Table A3 we test whether the rigidity of 0-ending prices changed after January 2008 at convenience stores with and without tills. To do so, we add to the regression a dummy for the period following January 2008, and its interactions with 0-ending prices, 9-ending prices, transaction inconvenience, and stores without tills.

We find that before 2008, 0-ending prices were more rigid than other prices at convenience stores ($\beta = -0.056, p < 0.01$). We also find no statistically significant differences in the rigidity of 0-ending prices at convenience stores with and without a till ($\beta = -0.062, p > 0.55$).

After 2008, we find that 0-ending prices became more rigid at convenience stores with a till than before 2008 ($\beta = -0.023, p < 0.01$). The difference between stores with and without tills is not statistically significant ($\beta = 0.164, p > 0.10$). Thus, after 2008, 0-ending prices were more rigid than before 2008 at convenience stores with a till, and as rigid as they were before 2008 at convenience stores without a till ($F = 2.0, p > 0.15$).

In summary, our results suggest that convenience played some role in the use of 0-ending prices at convenience stores. Convenience stores where the retailers had a greater incentive to use 0-ending prices to reduce computation effort and time used more 0-ending prices than stores that had less incentive to do so. Nevertheless, the share of 0-ending prices was over 60% also in stores that had little incentive to use 0-ending prices for reducing the computation effort. We therefore conclude that computation effort was not the only reason for using 0-ending prices, and perhaps not the most important also.

Table A1. Probability of a price change at convenience stores and superstores, Israel

| | **(1)** | **(2)** | **(3)** | **(4)** | **(5)** | **(6)** | **(7)** | (8) |
|---|---|---|---|---|---|---|---|---|
| 0-Ending | 0.010***<br>(0.000) | 0.089***<br>(0.001) | 0.002***<br>(0.000) | 0.008***<br>(0.000) | 0.001<br>(0.002) | −0.056***<br>(0.000) | −0.044***<br>(0.002) | −0.132***<br>(0.010) |
| 9-Ending | −0.045***<br>(0.000) | −0.038***<br>(0.001) | −0.054***<br>(0.000) | −0.047***<br>(0.000) | −0.201***<br>(0.002) | −0.018***<br>(0.000) | −0.110***<br>(0.002) | −0.141***<br>(0.012) |
| Transaction-Inconvenience | 0.017***<br>(0.000) | 0.022***<br>(0.000) | | 0.009***<br>(0.000) | 0.035***<br>(0.000) | 0.014***<br>(0.000) | 0.001***<br>(0.000) | −0.013***<br>(0.000) |
| Convenience Store | −0.043***<br>(0.000) | −0.050***<br>(0.001) | −0.037***<br>(0.000) | −0.053***<br>(0.000) | −0.234***<br>(0.000) | | | |
| Convenience× 0-Ending | −0.063***<br>(0.000) | −0.113***<br>(0.002) | −0.071***<br>(0.000) | −0.053***<br>(0.000) | −0.218***<br>(0.000) | | | |
| Convenience× 9-Ending | 0.030***<br>(0.000) | 0.003***<br>(0.001) | 0.018***<br>(0.000) | 0.036***<br>(0.000) | 0.121***<br>(0.000) | | | |
| Convenience× Transaction-Inconvenience | −0.003***<br>(0.000) | −0.025***<br>(0.000) | | 0.001***<br>(0.000) | 0.005***<br>(0.000) | | | |
| Price Level | 0.001***<br>(0.000) | 0.012***<br>(0.000) | 0.000***<br>(0.000) | 0.000***<br>(0.000) | 0.000***<br>(0.000) | 0.000***<br>(0.000) | 0.000***<br>(0.000) | 0.020***<br>(0.000) |
| Log of the Population | 0.005***<br>(0.000) | 0.012***<br>(0.000) | 0.010***<br>(0.000) | 0.005***<br>(0.000) | 0.015***<br>(0.000) | 0.005***<br>(0.000) | −0.016***<br>(0.000) | 0.014***<br>(0.001) |
| Socio-Economic Score | −0.004***<br>(0.000) | −0.006***<br>(0.000) | −0.003***<br>(0.000) | −0.003***<br>(0.000) | −0.013***<br>(0.000) | −0.003***<br>(0.000) | −0.009***<br>(0.000) | 0.002***<br>(0.000) |
| Share of Women | 0.034***<br>(0.000) | 0.022***<br>(0.000) | 0.083***<br>(0.000) | 0.032***<br>(0.000) | 0.112***<br>(0.000) | 0.034***<br>(0.002) | 1.159<br>(0.839) | −10.803<br>(165.526) |
| Log of the Distance from Tel-Aviv (in km) | −0.007***<br>(0.000) | −0.014***<br>(0.000) | −0.011***<br>(0.000) | −0.008***<br>(0.000) | −0.033***<br>(0.000) | −0.005***<br>(0.000) | −0.013***<br>(0.000) | −0.070***<br>(0.001) |
| Share of Minority Groups | 0.000***<br>(0.000) | 0.000***<br>(0.000) | 0.000***<br>(0.000) | 0.000***<br>(0.000) | 0.000***<br>(0.000) | 0.000***<br>(0.000) | 0.001***<br>(0.000) | 0.001***<br>(0.000) |
| Sale Dummy | 0.440***<br>(0.000) | 0.373***<br>(0.000) | 0.534***<br>(0.000) | | 4.351***<br>(0.081) | | | |
| 2-Ending | | | | | | 0.014***<br>(0.000) | 0.024***<br>(0.002) | |
| 3-Ending | | | | | | 0.041***<br>(0.000) | −0.011***<br>(0.003) | |
| 4-Ending | | | | | | 0.033***<br>(0.000) | 0.035***<br>(0.003) | |
| 5-Ending | | | | | | −0.007***<br>(0.000) | −0.090***<br>(0.002) | |
| 6-Ending | | | | | | −0.006***<br>(0.000) | −0.032***<br>(0.002) | |
| 7-Ending | | | | | | −0.009***<br>(0.000) | −0.084***<br>(0.002) | |
| 8-Ending | | | | | | 0.026***<br>(0.000) | 0.002**<br>(0.001) | |
| Constant | 0.181***<br>(0.006) | −0.16<br>(0.012) | 0.152***<br>(0.008) | 0.215***<br>(0.005) | −0.683***<br>(0.064) | 0.128***<br>(0.006) | 0.079<br>(0.196) | 5.750<br>(40.593) |
| $R^2$ | 0.341 | 0.255 | 0.323 | 0.282 | | 0.341 | 0.136 | 0.056 |
| $\chi^2$ | | | | | 50,415.7 | | | |
| Number of Observations | 527,633 | 130,505 | 85,837 | 514,682 | 564,742 | 442,811 | 121,931 | 3,585 |

Notes

The table presents the results of estimating regressions of the probability of a price change. Columns 1–4 and 6–8 report the results of estimating linear probability regressions. Column 5 reports the results of estimating a probit regression. The independent variable in all regressions is a dummy that equals 1 if the price has changed in a given month. The independent variables are: 0-Ending – a dummy for 0 ending prices, 9-Ending – a dummy for 9-ending prices, Transaction-Inconvenience – the minimum number of coins necessary for paying a price in cash, Convenience Store – an store that is defined as a convenience store/store, Price Level – the price in month $t$ – 1 rounded to the nearest NIS, Log of the Population – the log of the number of inhabitants in the town where the store is located, Socio-Economic Score – the socio-economic score of the town where the store is located, as reported by the Israel's Central Bureau of Statistics (CBS), Share of Women – the share of women in the town where the store is located, Log of the Distance from Tel-Aviv (in km) – the log of the distance of the town where the store is located from Tel-Aviv, Share of Minority Groups – the share of minority groups in the town where the store is located, Sale Dummy – a dummy for sale prices in month $t$ – 1, 2-Ending − a dummy, which equals 1 if the previous price was 2-ending, 3-Ending − a dummy, which equals 1 if the previous price was 3-ending, 4-Ending − a dummy, which equals 1 if the previous price was 4-ending, 5-Ending − a dummy, which equals 1 if the previous price was 5-ending, 6-Ending − a dummy, which equals 1 if the previous price was 6-ending, 7-Ending − a dummy, which equals 1 if the previous price was 7-ending, 8-Ending − a dummy, which equals 1 if the previous price was 8-ending. All regressions also include controls for product categories, the six regions of Israel, and for the quarter in which the observation was taken. Column 1 includes observations on prices ≤ 200 NIS. Column 2 includes observations on prices ≤ 4 NIS. Column 3 includes observations on prices that can be paid using a single coin or a single note. Column 4 includes observations on regular prices. Column 5 includes all observations. Column 6 includes all observations from convenience stores. Column 7 includes all observations from superstores. Column 8 reports the results of estimating the regression using observations only from stores that the CBS flags as convenience stores. Standard errors are clustered at the level of the store where the observation was taken.

* $p$ < 10%, ** $p$ < 5%, *** $p$ < 1%

Table A2. Testing the significance of the coefficient of 0-endings vs. other endings

| 0-Ending vs. | In Convenience Stores | In Superstores |
|---|---|---|
| 1-Ending | 58.71*** | 1.22 |
| 2-Ending | 46.50*** | 3.83* |
| 3-Ending | 71.31*** | 0.88 |
| 4-Ending | 106.05*** | 3.45* |
| 5-Ending | 24.87*** | 9.55*** |
| 6-Ending | 19.07*** | 0.19 |
| 7-Ending | 23.64*** | 2.08 |
| 8-Ending | 74.41*** | 2.92* |
| 9-Ending | 31.22*** | 26.52*** |

Note

The figures in the table are the $F$ test statistic values for comparing the coefficient of 0-ending prices with the coefficients of 1-ending, 2-ending… 9-ending prices, based on the results of the regressions reported in columns 6 and 7 of Table A1.

* $p < 10\%$, ** $p < 5\%$, *** $p < 1\%$

Table A3. Price rigidity at stores with and without a till

| | (1) | (2) |
|---|---|---|
| 0-Ending | -0.062***<br>(0.007) | -0.056***<br>(0.008) |
| 9-Ending | -0.035***<br>(0.012) | -0.011<br>(0.015) |
| Transaction-Inconvenience | 0.012***<br>(0.002) | 0.014***<br>(0.002) |
| Stores with no till dummy | 0.040<br>(0.081) | 0.073<br>(0.104) |
| Stores with no till dummy × 0-Ending | -0.026<br>(0.080) | -0.062<br>(0.103) |
| Stores with no till dummy × 9-Ending | -0.070<br>(0.081) | -0.113<br>(0.104) |
| Stores with no till dummy × Transaction-Inconvenience | 0.010***<br>(0.003) | 0.012***<br>(0.004) |
| Price Level | 0.000***<br>(0.000) | 0.000***<br>(0.000) |
| Log of the Population | 0.004<br>(0.005) | 0.004<br>(0.005) |
| Socio-Economic Score | -0.003<br>(0.002) | -0.003<br>(0.002) |
| Share of Women | 0.027<br>(0.040) | 0.026<br>(0.039) |
| Log of the Distance from Tel-Aviv (in km) | -0.004<br>(0.007) | -0.004<br>(0.007) |
| Share of Minority Groups | -0.000<br>(0.000) | -0.000<br>(0.000) |
| Sale dummy | 0.396***<br>(0.008) | 0.396***<br>(0.008) |
| Post-2008 dummy | | 0.001<br>(0.016) |
| Post-2008 dummy × 0-ending | | -0.023**<br>(0.008) |
| Post-2008 dummy × 9-ending | | -0.044***<br>(0.010) |
| Post-2008 dummy × Transaction inconvenience | | -0.004***<br>(0.001) |
| Post-2008 dummy × Stores with no till dummy | | -0.147<br>(0.099) |
| Post-2008 dummy × Stores with no till dummy × 0-ending | | 0.164<br>(0.100) |
| Post-2008 dummy × Stores with no till dummy × 9-ending | | 0.169*<br>(0.101) |
| Post-2008 dummy × Stores with no till dummy × Transaction inconvenience | | -0.008**<br>(0.004) |
| Constant | 0.146**<br>(0.073) | 0.141**<br>(0.074) |
| $R^2$ | 0.343 | 0.387 |
| Number of Observations | 442,811 | 442,811 |

Notes

The table presents the results of estimating regressions of the probability of a price change. The independent variable in all regressions is a dummy that equals 1 if the price has changed in a given month. The independent variables are: 0-Ending – a dummy for 0 ending prices, 9-Ending – a dummy for 9-ending prices, Transaction-Inconvenience – the minimum number of coins necessary for paying a price in cash, Stores with no till dummy – A dummy that equals 1 if the store is either an open market stall or a kiosk, Price Level – the price in month $t$ – 1 rounded to the nearest NIS, Log of the Population – the log of the number of inhabitants in the town where the store is located, Socio-Economic Score – the socio-economic score of the town where the store is located, as reported by the Israel's Central Bureau of Statistics (CBS), Share of Women – the share of women in the town where the store is located, Log of the Distance from Tel-Aviv (in km) – the log of the distance of the town where the store is located from Tel-Aviv, Share of Minority Groups – the share of minority groups in the town where the store is located, Sale Dummy – a dummy for sale prices in month $t$ – 1, post-2008 dummy – a dummy that equal 1 for observations after January 2008. Standard errors are clustered at the level of the store where the observation was taken.

* $p < 10\%$, ** $p < 5\%$, *** $p < 1\%$

**Appendix B. Robustness tests with the Dominick's data: estimating price rigidity**

In the paper, we report that for the goods in the front-end-candies category at Dominick's, 0-ending prices are common, and relatively rigid. Below, we show that this phenomenon is unique to the front-end-candies category: in other product categories at Dominick's, 0-endings are not particularly common or rigid.

Figure B1 depicts the distribution of the price endings in each of the 29 product categories at Dominick's. Table B1 complements it by showing the share of 0-ending prices in each category. It also provides the results of a Wilcoxon rank-sum test for comparing the share of 0-ending prices in the front-end-candies category with the shares in each of the other 28 categories.

Table B1 shows that the share of 0-ending prices in the front-end-candies category, 24.15 percent, is exceptionally high for Dominick's. The category with the next highest share of 0-ending prices is frozen dinners, with 13.12 percent. Thus, the share of 0-ending prices in the front-end-candies category is 84 percent higher than the share of 0-ending prices in the category with the next highest share of 0-ending prices, and more than 5 times the cross-category average of 4.74 percent.

Table B2 reports the results of estimating regressions of the likelihood of price changes in each of Dominick's 29 product categories. These regressions are equivalent to the regressions reported in Column 2 of Table 4 in the paper. The dependent variable is a dummy that equals 1 if the price has changed, and 0 otherwise. The independent variables include a dummy for 0-ending prices, a dummy for 9-ending prices, a transaction-inconvenience score that equals the minimum number of coins needed for paying the price, a dummy for quarter-multiple that equals 1 if the price ends in 25 or 75, the price level, calculated as the previous price rounded to the nearest dollar, the absolute value of changes in the wholesale price, and a dummy for sale prices, based on a 4-week sales filter.[4] The regression also includes sub-cateogries×stores×weeks fixed effects. We cluster the standard errors at the sub-categories×stores×weeks level.

[4] We exclude outlier observations, which we define conservatively as a change in the wholesale price in excess of 150 percent. This results in dropping 578,198 observations, which comprise about 0.6% of the total number of observations.

To save space, we report only the coefficients of the main variables of interest: The 0-ending price dummy, the 9-ending price dummy, the quarter-multiple dummy, and the transaction-inconvenience index. From the table, we can see that in all 29-categories, 9-ending prices are more rigid than other prices. The coefficient of 0-ending prices, however, is negative in only 5 product-categories: front end candies, frozen entrees, grooming products, soft drinks, and toothbrushes.

Table B3 reports the results of the same regressions, except that now we restrict the sample to regular prices only by removing sale and bounce back prices, based on a 4-week sales filter. Similar to the above, 9-ending prices are more rigid than other prices in all 29 product categories. 0-ending prices, however, are more rigid than other prices in only 2 product categories: front-end-candies and toothbrushes.

The results of these analyses, therefore, do not support the hypothesis that 0-endings are used as a signal of quality, because then 0-ending prices would be more rigid than other prices in other categories as well (Stiving 2000).

Thus, neither the popularity nor the rigidity of 0-ending prices is common at Dominick's stores. We believe that the special nature of the front-end-candies category, where shoppers make spontaneous decisions after considering perhaps only 1 or 2 products, can explain why 0-ending prices are both common and rigid in this category.

Table B4 contains the results of several robustness tests of the results we report in the paper for the front-end-candies category. In column 1 we use the same specification as in Tables B2 and B3. However, here we use all observations, including those with wholesale price changes in excess of 150%. Including these outlier observations reduces the effect of the wholesale price changes ($\beta = 0.000, p < 0.01$), but the effect of 0-ending prices remains negative ($\beta = -0.094, p < 0.01$).

In column 2 we use Dominick's sale price indicator rather than our sales filter to identify sales. As Peltzman (2000) notes, the Dominick's sales indicator was not set on a regular basis and, consequently, it is unreliable. Nevertheless, when we use it for testing the robustness of our results, we find that the coefficient of 0-ending prices is negative and significant ($\beta = -0.146, p < 0.01$).

In column 3 we restrict the data to observations on regular prices by removing observations on sale and bounce-back prices, using Dominick's sale indicator. We find that 0-ending prices are more rigid than other prices in this sample as well ($\beta = -0.114, p < 0.01$).

In column 4, we restrict the sample to sale prices only, again using Dominick's sales indicator. The coefficient of 0-ending prices is positive and significant ($\beta = 0.016, p < 0.05$). We also find that 9-ending ($\beta = 0.142, p < 0.01$) and 5-ending prices ($\beta = 0.235, p < 0.01$) are more likely to change than other prices. Thus, when we look at sale prices, it does not seem that the price ending affects the price rigidity. However, these results must be interpreted with caution, because as noted above the Dominick's sale indicator was not set on a regular basis. Consequently, there are many sale prices that are not accounted for in this regression.

In column 5, we add dummy variables for each possible ending, using the ending of 1 as a control. We find that the prices that are the most rigid are those that end with 9, 5, and 0. In other words, the most rigid prices are those that end with either 9, or the prices with "round" endings (i.e., 0 and 5). In Table B5, we show that the differences between the coefficient of 0-ending prices and the coefficients of the prices with other endings are statistically significant. These results confirm that 0, 9 and 5 are the most rigid endings at Dominick's front end candies category.

In column 6, we use only the data points where the price was the same over two subsequent observations. Our data comes from scanner data, and therefore, prices are calculated as revenue over quantity sold. Thus, there is a chance that some price changes in the dataset are due to the method by which prices are calculated (Strulov-Shlain, 2022, Cambell and Eden, 2014). We may get a "spurious" price change, for example, if some transactions were not properly recorded, or if some consumers used bonus coupons. By removing prices that last only one week, we  reduce the risk that such measurement errors affect our results, because it is unlikely that there will be measurement errors two weeks in a row that give the exact same price (Strulov-Shlain, 2022). The results show that after excluding all the prices that lasted only one week, 0-ending prices are still more rigid than other prices ($\beta = -0.153, p < 0.01$).

Thus, including all observations, using Dominick's sales indicator instead of a sales filter, or removing observations on prices that last one week, does not change the results we report in the paper. In the front-end-candies category, 0-ending prices are more rigid than other prices regardless of whether they are regular or sale prices.

As a final robustness check, we examine if our results hold when we include data from before January 1991. In the paper, we remove data collected before 1991 because in the earlier part of the data, Dominick's participated in a pricing experiment in cooporation with the faculty of the Booth School of Business at the University of Chicago, and this might have affected the outcomes. Table B6 gives the results when we use the full set of data.

In column 1, the only independent variables are the 0-ending dummy, the 9-ending dummy and the transaction-inconvenience score. We find that 0-endings ($\beta = -0.018, p < 0.01$) and 9-endings ($\beta = -0.135, p < 0.00$) reduce the likelihood of a price change. An increase in the transaction-inconvenience score increases the likelihood of a price change ($\beta = 0.041, p < 0.01$).

In column 2, we add the following controls: a dummy for quarter-multiple, the price level, the absolute value of changes in the wholesale price, and a dummy for sale prices, all defined as above.[5] The regression also includes sub-cateogries×stores×weeks fixed effects.

We find that the effects of 0-ending prices ($\beta = -0.137, p < 0.01$) and 9-ending prices ($\beta = -0.166, p < 0.01$) remain negative and statistically significant. The effect of the transaction-inconvenience score is still positive and statistically significant ($\beta = 0.016, p < 0.01$).

Thus, using the full dataset does not change any of our main results. 0-ending prices, as well as 9-ending prices, are more rigid than other ending prices.

[5] We exclude outlier observations, which we define conservatively as a change in the wholesale price in excess of 150 percent. When we use the full set of data, this results in dropping 66,232 observations, about 1.5% of the total number of observations.

Figure B1. Frequency distribution of the last digit of the retail prices at Dominick's, by product category, September 14, 1989–May 8, 1997

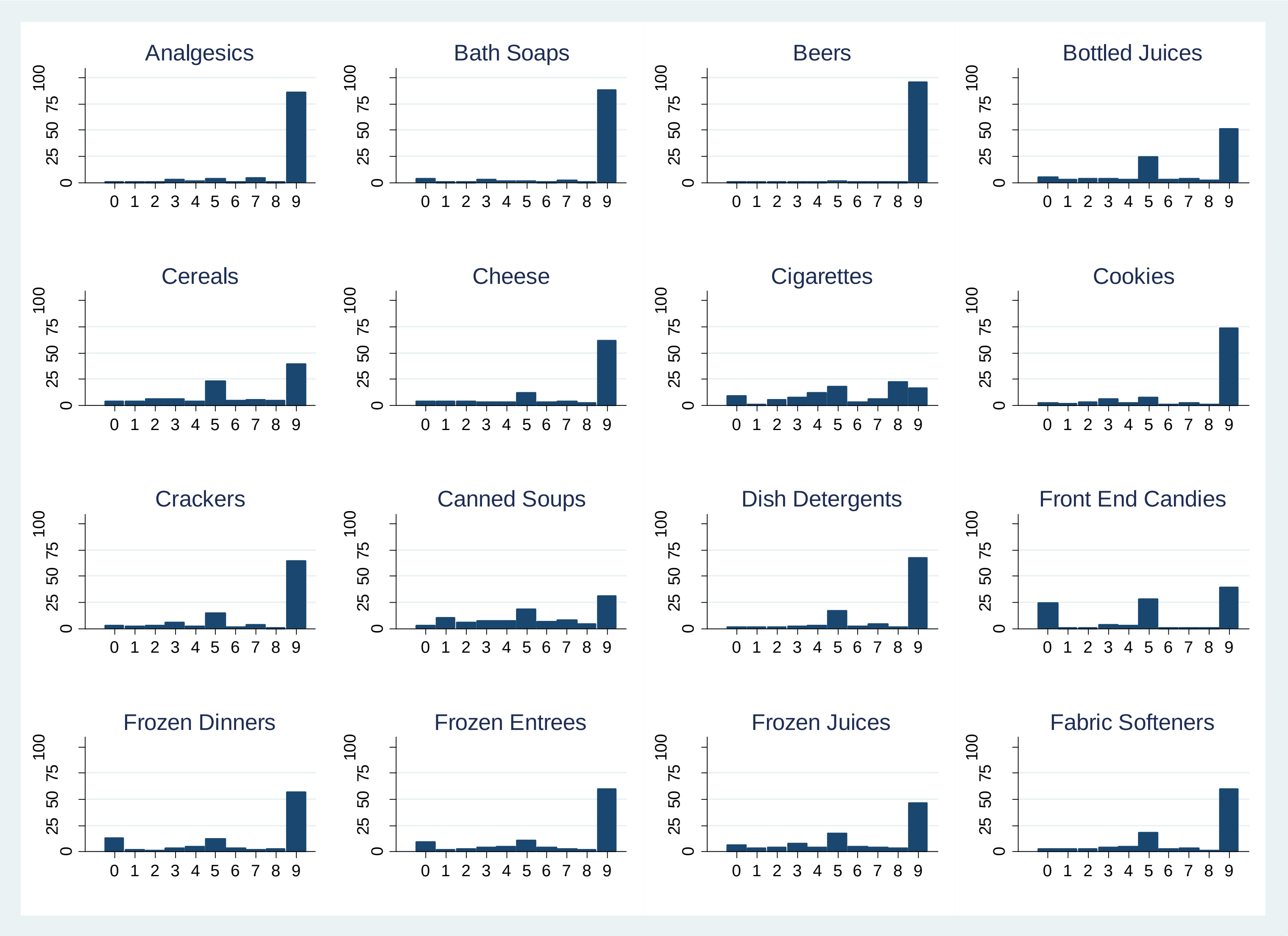

Figure B1. Frequency distribution of the last digit of the retail prices at Dominick's, by product categories, September 14, 1989–May 8, 1997 (Cont.)

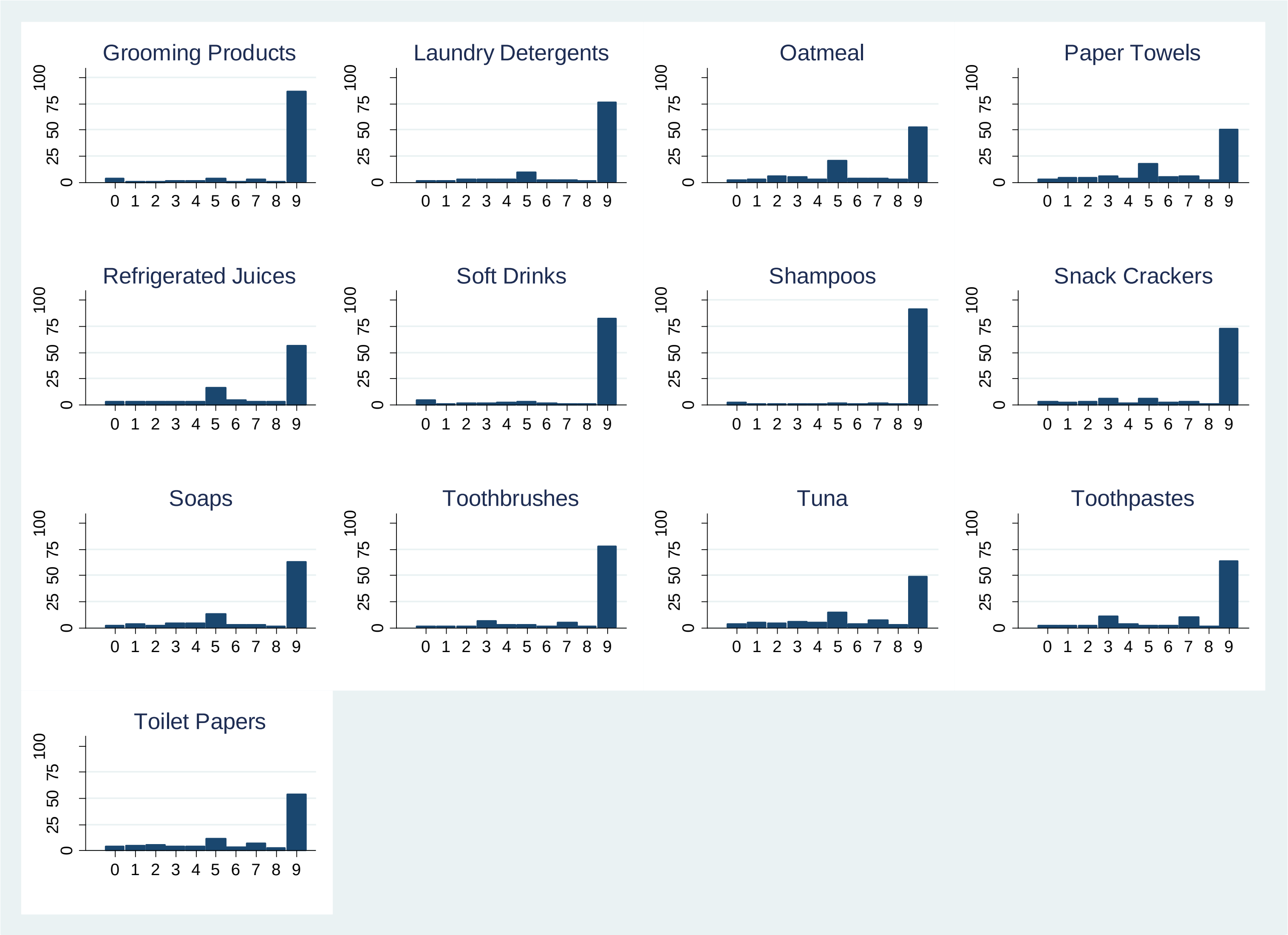

Table B1. Percentage of 0-ending prices at Dominick's, by product categories

| **Product Category** | **% 0-Ending Prices** | **No. of Observations** | **Wilcoxon Rank-Sum Test** |
|---|---|---|---|
| Analgesics | 0.39 | 3,040,172 | 906.61*** |
| Bath Soaps | 3.07 | 418,097 | 312.79*** |
| Beers | 0.39 | 1,966,148 | 740.85*** |
| Bottled Juices | 4.84 | 4,325,024 | 808.72*** |
| Canned Soups | 2.77 | 5,504,494 | 1023.43*** |
| Cereals | 4.12 | 4,707,776 | 876.66*** |
| Cheese | 4.15 | 6,752,328 | 1014.52*** |
| Cigarettes | 8.69 | 1,801,470 | 440.12*** |
| Cookies | 2.14 | 7,568,429 | 1243.23*** |
| Crackers | 2.92 | 2,228,269 | 687.34*** |
| Dish Detergents | 1.21 | 2,164,793 | 742.81*** |
| Fabric Softeners | 2.81 | 2,278,995 | 698.42*** |
| Front-End-Candies | 24.15 | 4,437,054 | N/A |
| Frozen Dinners | 13.12 | 1,654,053 | 296.33*** |
| Frozen Entrees | 9.32 | 7,172,075 | 687.77*** |
| Frozen Juices | 6.52 | 2,368,157 | 570.00*** |
| Grooming Products | 3.37 | 4,065,694 | 866.49*** |
| Laundry Detergents | 1.25 | 3,277,445 | 895.12*** |
| Oatmeal | 2.23 | 981,037 | 489.39*** |
| Paper Towels | 2.73 | 940,757 | 468.26*** |
| Refrigerated Juices | 3.28 | 2,166,755 | 665.55*** |
| Shampoos | 2.41 | 4,676,790 | 975.85*** |
| Snack Crackers | 2.88 | 3,487,565 | 837.26*** |
| Soaps | 2.06 | 1,835,196 | 659.75*** |
| Soft Drinks | 4.25 | 10,741,743 | 1231.64*** |
| Toilet Papers | 4.20 | 1,149,973 | 476.27*** |
| Toothbrushes | 1.62 | 1,839,536 | 675.47*** |
| Toothpastes | 2.39 | 2,981,532 | 804.89*** |
| Tuna | 3.46 | 2,382,983 | 687.24*** |
| Average or Total | 4.74 | 98,914,340 | |

<u>Notes</u>

The table presents the percentage of 0-ending prices in each of Dominick's 29 product categories. The final column reports the results of the $z$-statistic for the Wilcoxon rank sum test for comparing the share of 0-ending prices in the front-end-candies category with the share of 0-ending prices in each of the other 28 product categories.

* $p < 10\%$, ** $p < 5\%$, *** $p < 1\%$

Table B2. Price rigidity of the products at Dominick's, by product categories, all observations

| Category | Main Independent Variable | | | | |
|---|---|---|---|---|---|
| | **0-Ending Price** | **9-Ending Price** | **Quarter-Multiple Price** | **Transaction-Inconvenience** | ***N*** |
| Analgesics | 0.226*** (0.001) | −0.098*** (0.000) | −0.056*** (0.001) | −0.007*** (0.000) | 2,997,267 |
| Bath Soaps | 0.068*** (0.004) | −0.226*** (0.001) | 0.063*** (0.000) | −0.005*** (0.000) | 402,600 |
| Beers | 0.116*** (0.001) | −0.500*** (0.000) | −0.426*** (0.000) | 0.011*** (0.000) | 1,936,341 |
| Bottled Juices | 0.159*** (0.000) | −0.066*** (0.000) | −0.016*** (0.000) | 0.015*** (0.000) | 4,276,615 |
| Canned Soups | 0.183*** (0.000) | −0.026*** (0.000) | −0.027*** (0.000) | 0.005*** (0.000) | 5,450,234 |
| Cereals | 0.080*** (0.000) | −0.027*** (0.000) | −0.014*** (0.000) | −0.001*** (0.000) | 4,661,586 |
| Cheese | 0.141*** (0.000) | −0.186*** (0.000) | −0.073*** (0.001) | 0.012*** (0.000) | 6,696,191 |
| Cigarettes | 0.040*** (0.000) | −0.014*** (0.000) | −0.001*** (0.000) | 0.009*** (0.000) | 1,762,231 |
| Cookies | 0.122*** (0.000) | −0.168*** (0.000) | −0.008*** (0.000) | 0.012*** (0.000) | 7,471,949 |
| Crackers | 0.120*** (0.000) | −0.174*** (0.000) | −0.016*** (0.001) | 0.010*** (0.000) | 2,203,563 |
| Dish Detergents | 0.432*** (0.002) | −0.060*** (0.000) | −0.006*** (0.000) | 0.007*** (0.000) | 2,141,470 |
| Fabric Softeners | 0.115*** (0.002) | −0.046*** (0.000) | −0.059*** (0.000) | −0.001*** (0.000) | 2,252,077 |
| Front End Candies | −0.013*** (0.001) | −0.079*** (0.000) | 0.031*** (0.000) | 0.039*** (0.000) | 4,437,054 |
| Frozen Dinners | 0.074*** (0.004) | −0.249*** (0.000) | −0.148*** (0.002) | 0.006*** (0.000) | 1,623,448 |
| Frozen Entrees | −0.014*** (0.001) | −0.176*** (0.000) | −0.026*** (0.001) | 0.017*** (0.000) | 6,997,451 |
| Frozen Juices | 0.010*** (0.000) | −0.017*** (0.000) | −0.013*** (0.000) | −0.002*** (0.000) | 2,339,853 |
| Grooming Products | −0.015*** (0.003) | −0.245*** (0.000) | 0.098*** (0.000) | −0.002*** (0.000) | 3,974,487 |
| Laundry Detergents | 0.153*** (0.001) | −0.084*** (0.000) | −0.070*** (0.000) | −0.010*** (0.000) | 3,230,290 |
| Oatmeal | 0.158*** (0.001) | −0.069*** (0.000) | −0.025*** (0.000) | −0.002*** (0.000) | 970,697 |
| Paper Towels | 0.179*** (0.002) | −0.020*** (0.001) | 0.073*** (0.004) | 0.021*** (0.000) | 924,672 |
| Refrigerated Juices | 0.267*** (0.000) | −0.116*** (0.000) | −0.160*** (0.002) | −0.005*** (0.000) | 2,146,335 |
| Shampoos | 0.179*** (0.002) | −0.244*** (0.001) | −0.058*** (0.002) | −0.010*** (0.000) | 4,529,320 |
| Snack Crackers | 0.003*** (0.000) | −0.096*** (0.000) | −0.159*** (0.000) | 0.006*** (0.000) | 2,479,891 |
| Soaps | 0.271*** (0.001) | −0.107*** (0.000) | −0.012*** (0.001) | 0.004*** (0.000) | 1,807,790 |
| Soft Drinks | −0.008*** (0.002) | −0.286*** (0.001) | −0.139*** (0.005) | −0.004*** (0.000) | 10,376,206 |
| Toilet Papers | 0.256***<br>(0.002) | −0.046***<br>(0.000) | −0.057***<br>(0.001) | −0.004***<br>(0.000) | 1,134,801 |
| Toothbrushes | −0.009***<br>(0.002) | −0.082***<br>(0.001) | −0.002***<br>(0.000) | −0.022***<br>(0.000) | 1,805,772 |
| Toothpastes | 0.091***<br>(0.002) | −0.027***<br>(0.000) | 0.158***<br>(0.002) | −0.022***<br>(0.000) | 2,939,561 |
| Tuna | 0.073***<br>(0.001) | −0.077***<br>(0.000) | −0.019***<br>(0.000) | 0.006***<br>(0.000) | 2,358,245 |

Notes

The table presents the results of a linear probability model regressions of the probability of a price change. The dependent variable is a dummy which equals 1 if the price of good *i* at store *j* changed on week *t*. The independent variables are: 0-Ending − a dummy which equals 1 if the previous price was 0-ending, 9-Ending − a dummy which equals 1 if the previous price was 9-ending, Quarter-multiple − a dummy which equals 1 if the previous price ended in a multiple of a quarter (coin), and Transaction-Inconvenience Score – the minimum number of coins needed to pay the previous price. The regressions also include the following variables: Price Level – the previous price rounded to the nearest dollar, Sale Price – a dummy for the previous price being a sale price, and Absolute Value of the Percentage Change in the Wholesale Price – the absolute percentage change in the wholesale price. The regressions also include products×stores fixed effects. We exclude outlier observations, which we define as a change in the wholesale price in excess of 100 percent. This results in dropping 578,198 observations, about 0.6% of the total. Robust standard errors, clustered at the store-product level, are reported in parentheses.

* $p < 10\%$, ** $p < 5\%$, *** $p < 1\%$

Table B3. Price rigidity of the products at Dominick's, by product categories, regular prices

| Category | Main Independent Variable | | | | |
|---|---|---|---|---|---|
| | **0-Ending Price** | **9-Ending Price** | **Quarter-Multiple Price** | **Transaction-Inconvenience** | ***N*** |
| Analgesics | 0.223***<br>(0.000) | −0.099***<br>(0.000) | −0.056***<br>(0.000) | −0.006***<br>(0.000) | 2,821,654 |
| Bath Soaps | 0.053***<br>(0.000) | −0.216***<br>(0.000) | 0.137***<br>(0.000) | 0.002***<br>(0.000) | 382,932 |
| Beers | 0.123***<br>(0.000) | −0.536***<br>(0.000) | −0.424***<br>(0.000) | 0.015***<br>(0.000) | 1,495,058 |
| Bottled Juices | 0.146***<br>(0.000) | −0.098***<br>(0.000) | −0.020***<br>(0.000) | 0.024***<br>(0.000) | 3,455,004 |
| Canned Soups | 0.162***<br>(0.000) | −0.037***<br>(0.000) | −0.036***<br>(0.000) | 0.006***<br>(0.000) | 4,730,051 |
| Cereals | 0.053***<br>(0.000) | −0.037***<br>(0.000) | −0.014***<br>(0.000) | 0.004***<br>(0.000) | 4,161,630 |
| Cheese | 0.149***<br>(0.000) | −0.176***<br>(0.000) | −0.063***<br>(0.000) | 0.013***<br>(0.000) | 5,129,546 |
| Cigarettes | 0.048***<br>(0.000) | −0.020***<br>(0.000) | 0.000***<br>(0.000) | 0.012***<br>(0.000) | 1,750,716 |
| Cookies | 0.173***<br>(0.000) | −0.176***<br>(0.000) | −0.049***<br>(0.000) | 0.016***<br>(0.000) | 6,269,874 |
| Crackers | 0.145***<br>(0.000) | −0.168***<br>(0.000) | −0.061***<br>(0.000) | 0.015***<br>(0.000) | 1,794,652 |
| Dish Detergents | 0.411***<br>(0.000) | −0.081***<br>(0.000) | −0.022***<br>(0.000) | 0.009***<br>(0.000) | 1,850,794 |
| Fabric Softeners | 0.094***<br>(0.000) | −0.074***<br>(0.000) | −0.061***<br>(0.000) | 0.006***<br>(0.000) | 1,987,360 |
| Front End Candies | −0.005***<br>(0.000) | −0.056***<br>(0.000) | 0.017***<br>(0.000) | 0.030***<br>(0.000) | 4,008,259 |
| Frozen Dinners | 0.057***<br>(0.000) | −0.277***<br>(0.000) | −0.213***<br>(0.000) | 0.013***<br>(0.000) | 1,212,996 |
| Frozen Entrees | 0.013***<br>(0.000) | −0.150<br>(0.000) | −0.044***<br>(0.000) | 0.017***<br>(0.000) | 5,616,611 |
| Frozen Juices | 0.017***<br>(0.000) | −0.048***<br>(0.000) | −0.005***<br>(0.000) | 0.003***<br>(0.000) | 1,806,969 |
| Grooming Products | 0.019***<br>(0.000) | −0.247***<br>(0.000) | 0.084***<br>(0.000) | 0.001***<br>(0.000) | 3,569,479 |
| Laundry Detergents | 0.133***<br>(0.000) | −0.101***<br>(0.000) | −0.057***<br>(0.000) | −0.005***<br>(0.000) | 2,816,612 |
| Oatmeal | 0.192***<br>(0.000) | −0.072***<br>(0.000) | −0.020***<br>(0.000) | 0.009***<br>(0.000) | 847,425 |
| Paper Towels | 0.175***<br>(0.000) | −0.051***<br>(0.000) | 0.024***<br>(0.000) | 0.020***<br>(0.000) | 722,957 |
| Refrigerated Juices | 0.295***<br>(0.000) | −0.150***<br>(0.000) | −0.119***<br>(0.000) | 0.010***<br>(0.000) | 1,459,298 |
| Shampoos | 0.135***<br>(0.000) | −0.258***<br>(0.000) | −0.026***<br>(0.000) | −0.005***<br>(0.000) | 4,122,264 |
| Snack Crackers | 0.001<br>(0.006) | −0.096***<br>(0.002) | −0.159***<br>(0.006) | 0.005***<br>(0.001) | 2,468,519 |
| Soaps | 0.273***<br>(0.000) | −0.118***<br>(0.000) | −0.040***<br>(0.000) | 0.006***<br>(0.000) | 1,555,553 |
| Soft Drinks | 0.049***<br>(0.000) | −0.394***<br>(0.000) | −0.195***<br>(0.000) | −0.004***<br>(0.000) | 6,807,239 |
| Toilet Papers | 0.198***<br>(0.000) | −0.089***<br>(0.000) | −0.075***<br>(0.000) | 0.000***<br>(0.000) | 880,326 |
| Toothbrushes | −0.031***<br>(0.000) | −0.084***<br>(0.000) | −0.002***<br>(0.000) | −0.022***<br>(0.000) | 1,625,533 |
| Toothpastes | 0.116***<br>(0.000) | −0.050***<br>(0.0000 | 0.165***<br>(0.000) | −0.018***<br>(0.000) | 2,506,048 |
| Tuna | 0.108***<br>(0.000) | −0.083***<br>(0.000) | −0.018***<br>(0.000) | 0.012***<br>(0.000) | 2,081,953 |

Notes

The table presents the results of the estimation of a linear probability model regressions of the probability of a *regular* price change. To focus on regular prices, we exclude the observations on sale prices and on bounce back prices, using a 4-week sale filter to identify sale prices. The dependent variable is a dummy which equals 1 if the price of good *i* at store *j* changed on week *t*. The independent variables are: 0-Ending − a dummy which equals 1 if the previous price was 0-ending, 9-Ending − a dummy which equals 1 if the previous price was 9-ending, Quarter-multiple − a dummy which equals 1 if the previous price ended in a multiple of a quarter (coin), and Transaction-Inconvenience Score – the minimum number of coins needed to pay the previous price. The regressions also include the following variables: Price Level – the previous price rounded to the nearest dollar, Sale Price – a dummy for the previous price being a sale price, and Absolute Value of the Percentage Change in the Wholesale Price – the absolute percentage change in the wholesale price. The regressions also include products×stores fixed effects. We exclude outlier observations, which we define as a change in the wholesale price in excess of 100 percent. This results in dropping 578,198 observations, about 0.6% of the total. Robust standard errors, clustered at the store-product level, are reported in parentheses.

* $p < 10\%$, ** $p < 5\%$, *** $p < 1\%$

Table B4. Price rigidity of the products in the front-end-candies' category at Dominick's

| | **(1)** | **(2)** | **(3)** | **(4)** | **(5)** | **(6)** |
|---|---|---|---|---|---|---|
| 0-Ending | −0.094*** (0.003) | −0.146*** (0.003) | −0.114*** (0.004) | 0.016** (0.008) | −0.0448*** (0.007) | −0.153*** (0.005) |
| 9-Ending | −0.160*** (0.003) | −0.188*** (0.003) | −0.157*** (0.004) | 0.142*** (0.006) | −0.487*** (0.007) | −0.163*** (0.004) |
| Transaction-Inconvenience Score | 0.021*** (0.001) | 0.019*** (0.001) | 0.017*** (0.001) | 0.004 (0.001) | 0.022*** (0.001) | −0.008*** (0.0010 |
| Quarter-multiple | −0.011*** (0.001) | −0.014*** (0.002) | 0.020*** (0.001) | 0.147*** (0.005) | −0.003** (0.001) | −0.030*** (0.001) |
| Price Level | −0.047*** (0.002) | −0.071*** (0.001) | −0.048*** (0.001) | 0.047*** (0.004) | −0.040*** (0.002) | −0.010*** (0.010) |
| Absolute Value of the % Change in the Wholesale Price | 0.000*** (0.000) | 0.011*** (0.000) | 0.009*** (0.000) | 0.008*** (0.000) | 0.011*** (0.000) | 0.010*** (0.000) |
| 5-Ending | −0.133*** (0.002) | −0.172*** (0.003) | −0.147*** (0.004) | 0.235*** (0.004) | −0.466*** (0.006) | −0.161*** (0.004) |
| Sale Price (Sales Filter) | 0.495*** (0.002) | | | | 0.391*** (0.003) | 0.387*** (0.002) |
| Sale Price (Dominick's Sales Dummy) | | 0.263*** (0.003) | | | | |
| 2-Ending | | | | | −0.304*** (0.007) | |
| 3-Ending | | | | | −0.428*** (0.008) | |
| 4-Ending | | | | | −0.374*** (0.008) | |
| 6-Ending | | | | | −0.122*** (0.009) | |
| 7-Ending | | | | | −0.015* (0.009) | |
| 8-Ending | | | | | −0.114*** (0.005) | |
| Constant | 0.242*** (0.003) | 0.226*** (0.003) | 0.167*** (0.004) | 0.310*** (0.006) | 0.489*** (0.006) | 0.204*** (0.006) |
| $R^2$ | 0.219 | 0.286 | 0.118 | 0.092 | 0.378 | 0.347 |
| *N* | 3,708,902 | 3,686,663 | 3,2143,57 | 327,180 | 3,686,663 | 3,483,539 |

<u>Notes</u>

The table presents the results of estimating a linear probability model regression of the probability of a price change. The dependent variable is a dummy, which equals 1 if the price of good *i* at store *j* changed on week *t*. The independent variables are: 0-Ending − a dummy which equals 1 if the previous price was 0-ending, 9-Ending − a dummy which equals 1 if the previous price was 9-ending, Quarter-multiple − a dummy which equals 1 if the previous price ended in 25 or 75, Transaction-Inconvenience Score – the minimum number of coins needed to pay the previous price, price level – the previous price rounded to the nearest dollar, Absolute Value of the Percentage Change in the Wholesale Price – the absolute percentage change in the wholesale price, Sale Price (Sales Filter) – a dummy for the previous price being a sale price based on a 4 week sale filter, Sale Price (Dominick's Sales Dummy) – a dummy for the previous price being a sale price, based on Dominick's sales indicator, 2-Ending − a dummy which equals 1 if the previous price was 2-ending, 3-Ending − a dummy which equals 1 if the previous price was 3-ending, 4-Ending − a dummy which equals 1 if the previous price was 4-ending, 5-Ending − a dummy which equals 1 if the previous price was 5-ending, 6-Ending − a dummy which equals 1 if the previous price was 6-ending, 7-Ending − a dummy which equals 1 if the previous price was 7-ending, and 8-Ending − a dummy which equals 1 if the previous price was 8-ending. The regressions also include products×stores fixed effects. The first and second columns contain all observations. In column 1, we estimate the regression using all observations, including those with changes in the wholesale price above 150%. In column 2, we estimate the regressions using Dominick's sales dummy instead of the sale filter to identify sales. In column 3, we estimate the regression using a sample of regular prices, by removing all observations that the Dominick's sales dummy identifies as sale prices, and the bounce back prices. In column 4, we estimate the regression using a sample of sale prices, by including only the prices that the Dominick's sales dummy identifies as sale prices. In column 5, we add dummy variables for each possible ending, using the ending 1 as a control. In column 6, we use only the data points where the price was the same over two subsequent observations. Robust standard errors, clustered at the store-product level, are reported in parentheses.
* $p < 10\%$, ** $p < 5\%$, *** $p < 1\%$

Table B5. Testing the significance of the coefficient of 0-ending vs. other endings

| **0-Ending vs.** | **Test-statistic** |
|---|---|
| 1-Ending | 4658.3*** |
| 2-Ending | 636.5*** |
| 3-Ending | 89.1*** |
| 4-Ending | 430.9*** |
| 5-Ending | 475.6*** |
| 6-Ending | 7693.7*** |
| 7-Ending | 10773.0*** |
| 8-Ending | 2741.2*** |
| 9-Ending | 1055.8*** |

Note

The figures in the table are the $F$ test statistic values for comparing the coefficient of 0-ending prices with the coefficients of 1-ending, 2-ending, …, 9-ending prices, based on the results of the regressions reported in column 5 of Table F4.

* $p < 10\%$, ** $p < 5\%$, *** $p < 1\%$

Table B6. Price rigidity of the products in the front-end-candies category at Dominick's, full dataset with all observations

| | **(1)** | **(2)** |
|---|---|---|
| 0-Ending | -0.018***<br>(0.001) | -0.137***<br>(0.003) |
| 9-Ending | -0.135***<br>(0.002) | -0.166***<br>(0.003) |
| Transaction-Inconvenience Score | 0.041***<br>(0.001) | 0.016***<br>(0.001) |
| Quarter-multiple | | -0.008***<br>(0.001) |
| Price Level | | -0.040***<br>(0.001) |
| Absolute Value of the % Change in the Wholesale Price | | 0.012***<br>(0.000) |
| 5-Ending | | -0.151***<br>(0.002) |
| Sale Price (Sales Filter) | | 0.416***<br>(0.003) |
| Constant | 0.075***<br>(0.002) | 0.185***<br>(0.002) |
| $R^2$ | 0.031 | 0.361 |
| $N$ | 4,402,665 | 4,402,665 |

Notes

The table presents the results of estimating a linear probability model regression of the probability of a price change. The dependent variable is a dummy, which equals 1 if the price of good $i$ at store $j$ changed on week $t$. The independent variables are: 0-Ending − a dummy which equals 1 if the previous price was 0-ending, 9-Ending − a dummy which equals 1 if the previous price was 9-ending, Quarter-multiple − a dummy which equals 1 if the previous price ended in 25 or 75, Transaction-Inconvenience Score – the minimum number of coins needed to pay the previous price, price level – the previous price rounded to the nearest dollar, Absolute Value of the Percentage Change in the Wholesale Price – the absolute percentage change in the wholesale price, Sale Price (Sales Filter) – a dummy for the previous price being a sale price based on a 4 week sale filter. The regressions also include sub-categories×stores×weeks fixed effects. The regressions use the full set of data, including observations from before 1991. Robust standard errors, clustered at the store-product level, are reported in parentheses. * $p < 10\%$, ** $p < 5\%$, *** $p < 1\%$

**Appendix C. Robustness tests with Dominick's data: estimating demand**

Table C1 contains the results of several robustness tests for the estimation of demand equations for front-end-candies. In panel 1 of Table 3 in the paper, we report the estimation results of regressions of the product-level demand equations: the dependent variable is the log of the quantity of product *q* sold at store *s*, in week *t*. The independent variables include dummies for 0- and 9-ending prices, the transaction-inconvenience score, the log of the price, the log of the average price of other products in the same sub-category, a dummy for sale prices, the quantity sold in store *s* in week *t* – 1, and fixed effects of stores, years, and months. We add the average price of other products in the same subcategory as a control for competition. We add the year and month fixed effects to control for possible seasonality in demand (Butters et al., 2020). We also add fixed effects for holidays, which include Christmas, New Year, Presidents' Day, Easter, Memorial Day, 4$^{th}$ of July, Labor Day, Halloween, and Thanksgiving. To minimize the effect of endogeneity, we use the log of the average price in other stores as instrument for the price.

In panel A, we report the results when we remove the control for the quantity sold in store *s* in week *t* – 1. This has little effect on the coefficients of 0- and 9-ending prices. The mean effect of 0-endings is positive ($\bar{\beta} = 0.202$) and a large majority of the coefficients are positive: 55 are positive vs. 25 negative. The mean effect of 9-endings is negative ($\bar{\beta} = -0.014$), and a minority of the coefficients is positive: 32/80 coefficients are positive.

In the paper, we use the average price in other stores as instrument. This might not be ideal, because all the stores are located in one city, Chicago. Therefore, in panel B, we use the wholesale prices as instrument for the price. We find that the mean effect of 0-endings is positive ($\bar{\beta} = 0.135$). The mean effect of 9-ending is negative ($\bar{\beta} = -0.034$). Again, the majority of the 0-ending coefficients is positive (57/80), while a minority of the 9-ending coefficients is positive (31/80).

In the paper, we do not use observations collected before January 1991 because data from the earlier period could have been contaminated by pricing experiments that Dominick's has conducted. As a test of robustness, therefore, we estimate the regressions using the full set of data. The results are reported in panel C.

Table C1. Regressions of the effects of price endings on demand in the front-end-candies category: Dominick's, U.S.

| | Average | Quantity-adjusted average | Revenue-adjusted average | Num. of positive coefficients | Num. of Negative coefficients | Num. of positive and significant coefficients | Num. of negative and significant coefficients | No. of observations |
|---|---|---|---|---|---|---|---|---|
| | Panel A | | | | | | | |
| 0-ending price | 0.202 | 0.246 | 0.220 | 55 | 25 | 42 | 12 | 1,988,759 |
| 9-ending price | -0.014 | -0.038 | -0.030 | 32 | 45 | 26 | 39 | |
| Transaction-inconvenience score | 0.084 | 0.094 | 0.081 | 66 | 14 | 53 | 10 | |
| Log(price) | -0.258 | -0.299 | -0.378 | 20 | 60 | 15 | 55 | |
| | Panel B | | | | | | | |
| 0-ending price | 0.135 | 0.133 | 0.115 | 57 | 23 | 38 | 13 | 1,987,194 |
| 9-ending price | -0.034 | -0.065 | -0.053 | 31 | 46 | 25 | 39 | |
| Transaction-inconvenience score | 0.066 | 0.074 | 0.069 | 68 | 12 | 60 | 9 | |
| Log(price) | 0.031 | 0.069 | 0.036 | 54 | 26 | 41 | 16 | |
| | Panel C | | | | | | | |
| 0-ending price | 0.138 | 0.137 | 0.127 | 55 | 25 | 40 | 11 | 2,307,850 |
| 9-ending price | -0.062 | -0.085 | -0.068 | 28 | 49 | 18 | 40 | |
| Transaction-inconvenience score | 0.065 | 0.074 | 0.069 | 63 | 17 | 50 | 14 | |
| Log(price) | -0.109 | -0.110 | -0.214 | 42 | 38 | 37 | 33 | |

Notes

The table summarizes the estimation results of reduced form product-level demand equations for 80 products in the Dominick's front-end-candies category, a total of 80 regressions. In each regression, the dependent variable is the log of the quantity of product $q$ sold at store $s$, in week $t$. The independent variables include a dummy for 0-ending price (1 if the price ends with 0), a dummy for 9-ending prices (1 if the price ends with 9), the transaction-inconvenience score (the minimum number of coins needed to pay the previous price), the log of the price, the log of the average price of other products in the same sub-category, the quantity sold in the same store in the previous week, and fixed effects for years and months, for stores, and for holidays, which include Christmas, New Year, Presidents Day, Easter, Memorial Day, 4th of July, Labor Day, Halloween, and Thanksgiving. In panels A and C, the average price in other stores is used as an instrument for the price. In panel B, the log of the wholesale price is used as the instrument for the price. Column 1 gives the unadjusted average of the coefficients. Column 2 gives the quantity-adjusted average of the coefficients. Column 3 gives the revenue-adjusted average of the coefficients. Column 4 (5) gives the number of positive (negative) coefficients. Column 6 (7) gives the number of positive and significant (negative and significant) coefficients. Column 8 gives the total number of observations. In panel A we do not add the quantity in period $t-1$ to the regression. In panel C we use observations from before January 1991.

* $p < 10\%$, ** $p < 5\%$, *** $p < 1\%$

## Appendix D: Nielsen Data

The Nielsen Retail Scanner Data available from the Kilts Center for Marketing at the Booth School of Business at the University of Chicago, consists of weekly price, sales volume, and store merchandising data reported by participating retail store point-of-sale systems in all US markets. Depending on the year, data are included from approximately 30,000–50,000 participating grocery, drug, mass merchandising, and other retail stores. Products from all Nielsen-tracked categories are included in the data, such as food, non-food grocery items, health and beauty aids, and select general merchandise.

We use the data for 2019. The Nielsen dataset has the advantage that it is large and it is representative of the current US retail market. However, for our purpose, it has two major drawbacks. First, prices in the dataset are quantity-weighted weekly average prices. Therefore, if a price changed during the week, or if some consumers paid a price that is different than the posted price (e.g., when they used a coupon), then the prices reported in the dataset are likely to be different than the prices that consumers actually paid during the week.

Second, the observations are divided into broad departments, such as “dry grocery,” “alcoholic beverages,” “dairy,” etc. Each department is further divided into product groups. For example, the product groups “bread and baked goods,” “fruit – canned” and “juice, drinks – canned, bottled” belong to the dry grocery department. Product groups are further divided into narrowly defined product modules. For example, the “fruit – canned” product group includes 18 modules, including “canned fruit – cherries,” and “canned fruit – grapes.”

We therefore have information on narrowly defined product modules, but we do not have information on where the products were located in the store. Consequently, we do not know whether or not the goods are sold via the front-end display shelves.

To deal with the first problem, we follow Strulov-Shlain (2022) by including in our sample only prices that remain unchanged for at least two weeks in a row. As Strulov-Shlain (2022) shows, this greatly reduces the risk of using prices that were not actually paid by consumers. This, however, comes at a cost: we loose about 50% of the observations (Strulov-Shlain, 2022).

To deal with the second problem, we looked for modules containing products that are usually placed on the front-end display shelves. We found three such modules: chewing gum, bubble gum, and sugar-free chewing gum.

Because we drop a large number of observations, we do not estimate regressions of price rigidity. Indeed, such regressions would be suspect since we include in our sample only the prices that last at least two weeks in a row. However, we follow Strulov-Shlain (2022) and use the data for estimating demand.

When estimating demand, we have to take into account the fact that we use scanner data. Consequently, we have observations on a good only in weeks in which at least one unit was sold. If we ignore missing observations, therefore, we might overestimate the demand for goods that are bought in small numbers, because we only observe them in weeks in which at least one unit was sold. We therefore find for each product (UPC) the number of weeks in which it was sold across all stores, and remove the products in the first quartile in terms of the number of weeks for which we have information (Strulov-Shlain, 2022).[6]

Table 1 gives summary statistics of each product module. The average price in these modules is $1.45–$2.28. The average number of units sold per store per week (sales volume) is 4.10–4.31. The number of observations varies between 2,827,275 in the bubble gum product module and 51,571,803 in the sugar-free chewing gum module.

Figure 1 depicts, for each product module, the percentage of observations that ends with each of the possible 10 right-most digits. As may be expected, given that the data comes from large supermarkets, drugstores, etc., the most common price ending is 9. The round endings, 0 and 5, however, are also quite common. In particular, 0-ending prices compose 8.3%–17.8% of all price endings.

[6] We have data for one year, 52 weeks. 26 weeks is therefore half of the maximum possible number of observations.

We estimate demand using a simple demand equation:

$$Ln(sales\text{-}volume_{i,s,t}) = \alpha + \beta_1 \times Ln(price_{i,s,t}) + \beta_2 \times right\text{-}0_{i,s,t} + \beta_3 \times right\text{-}9_{i,s,t} + \beta_4 \times competitors` \, price_{i,s,t} + \beta_5 \times Christmas_t + \gamma_{i,s} + \varepsilon_{i,s,t}$$

where sales-volume is the number of units of product $i$ sold in store $s$ in week $t$. Price is the price of the product. To alleviate the possible problem of endogeneity, we use the average price in other stores in the same week as an instrument for the price. Right-0 is a dummy variable that equals 1 if the price is 0-ending and 0 otherwise. Right-9 is a dummy that equals 1 if the price is 9-ending and 0 otherwise. Competitors' price is the average price of other products in the product's module offered in week $t$ in store $s$. Christmas is a dummy variable that equals 1 in the week that includes December 25 and 0 otherwise. $\gamma$ are fixed effects for store-product combinations. $\varepsilon$ is a random error. We cluster the standard errors by store.

The estimation results are given in Table 2. We find that in all three product modules, the coefficients of 0-ending prices are positive. In the chewing gum and the sugar-free chewing gum product modules, the coefficients are statistically significant at the 1% level. In the bubble gum product module, which is significantly smaller than the other two modules, the coefficient is statistically significant at the 10% level.

The sizes of the coefficients also seem economically significant: 0-ending prices are correlated with an increase of 1.1%–5.3% in sales volumes. 9-ending prices, on the other hand, are correlated with a decrease in the sales volumes.

Thus, the results we find corroborate the results we report in the paper. 0-ending prices seem to be positively correlated with sales volumes for products that are usually sold via front-end display shelves. Further, the results we report here are likely to be a lower bound on the effect of 0-ending prices in the front-end candies' department, since it is very likely that at least some of the goods in our database were also sold in other departments. The results we report here are therefore likely to be a mixture of the effects of 0-ending prices on the demand for products sold via the front-end display shelves, and those on the demand for products sold elsewhere in the stores.

Table 1. Summary statistics: Chewing gum, bubble gum and sugar free chewing gum, AC Nielsen

| | Chewing gum | Bubble gum | Sugar free chewing gum |
|---|---|---|---|
| Average price | 1.552<br>(0.842) | 1.451<br>(0.557) | 2.284<br>(1.122) |
| Average sales volume | 4.101<br>(5.073) | 4.310<br>(6.056) | 4.254<br>(5.781) |
| % 0-ending | 8.3% | 17.8% | 10.6% |
| % 9-ending | 56.7% | 70.2% | 74.1% |
| Number of UPCs | 181 | 10 | 733 |
| Number of stores | 46,863 | 30,803 | 47,136 |
| *N* | 8,142,043 | 2,827,275 | 51,571,803 |

Notes

Summary statistics for products in the chewing gum, bubble gum and sugar free chewing gum product modules in the AC Nielsen's database. We remove observations that have prices that last exactly one week. We also remove observations in the lower quartile of the observations over all stores and weeks.

Table 2. Regressions of the effects of price endings on demand in several product modules: AC Nielsen

| | Chewing gum | Bubble gum | Sugar free chewing gum |
|---|---|---|---|
| Right-0 | 0.029***<br>(0.003) | 0.011*<br>(0.007) | 0.053***<br>(0.002) |
| Right-9 | −0.218***<br>(0.003) | −0.175***<br>(0.005) | −0.326***<br>(0.002) |
| Ln(price) | −0.997***<br>(0.016) | 0.237***<br>(0.019) | −0.204***<br>(0.004) |
| Ln(competitors-price) | 0.009***<br>(0.001) | −0.006***<br>(0.002) | −0.009***<br>(0.002) |
| Christmas | 0.001***<br>(0.002) | 0.114***<br>(0.003) | 0.020***<br>(0.001) |
| Constant | 1.359***<br>(0.005) | 1.044***<br>(0.008) | 1.361***<br>(0.004) |
| $R^2$ | 0.057 | 0.002 | 0.010 |
| $N$ | 8,190,525 | 2,827,275 | 51,571,803 |

<u>Notes</u>

Results of fixed effects regressions of the sales volumes. The dependent variable is the log of the number of units sold of good *i* in store *s* in week *t*. The independent variables are: right-0 – a dummy variable that equals 1 if the price is 0-ending and 0 otherwise. right-9 – a dummy variable that equals 1 if the price is 9-ending and 0 otherwise. Ln(price) – the log of the product's price. We use the average price of products in other stores to instrument for the price. Ln(competitors-price) – the average price of other products sold in the same week in the same store. Christmas – a dummy variable that equals 1 in the week that includes December 25 and 0 otherwise. The $R^2$ is the pseudo overall $R^2$. Standard errors, clustered at the store level are reported in parentheses.

* $p < 0.10$, ** $p < 0.05$, *** $p < 0.01$

—

Figure 1. Percentage of price endings

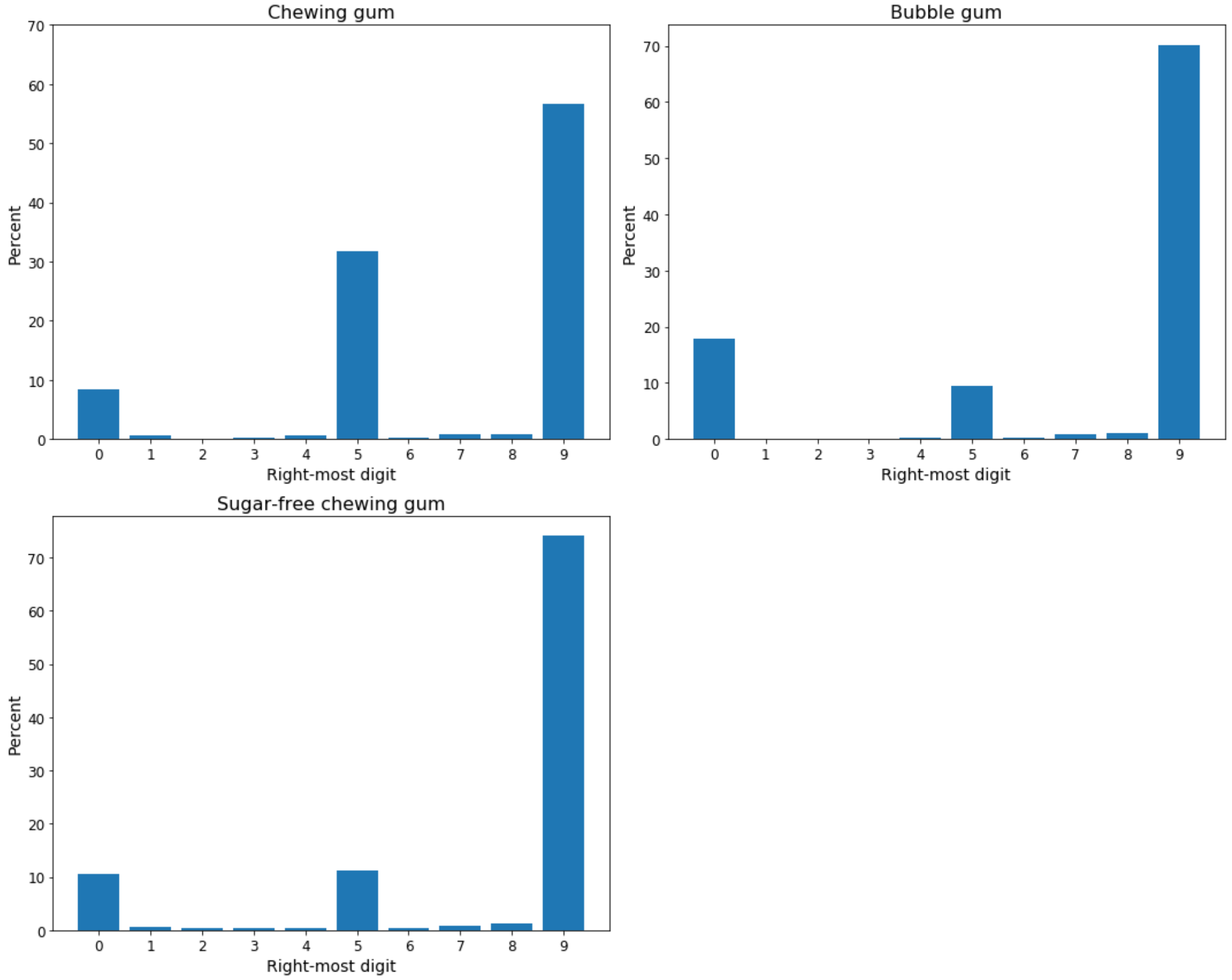